\documentclass{aa}
\usepackage{amssymb,amsmath}
\usepackage{graphicx,epstopdf}
\usepackage{natbib}
\usepackage{color}
\usepackage{xcolor}
\usepackage{media9}
\usepackage{dcolumn}
\usepackage{bm}
\usepackage{tikz}
\usepackage{filecontents}
\usepackage{arydshln}
\usepackage{longtable}
\usepackage{hyperref}
\usepackage{longtable}

\usepackage{ulem}

\usepackage[font=small,labelfont=bf,justification=justified]{subcaption}
\usepackage[font=small,labelfont=bf,justification=justified]{caption}
\graphicspath{{Images/}}
\definecolor{mygreen}{rgb}{0,0.6,0}
\definecolor{mygray}{rgb}{0.95,0.95,0.95}
\definecolor{mymauve}{rgb}{0.58,0,0.82}

\begin{document}

\title{Partially accreted crusts of neutron stars}
\author{L.Suleiman\inst{1,2} \and J.L. Zdunik \inst{1} \and P. Haensel \inst{1} \and  M. Fortin\inst{1}}

\institute{N. Copernicus Astronomical Center, Polish Academy of Sciences, 
Bartycka 18, PL-00-716 Warszawa, Poland
\and Laboratoire Univers et Th\'{e}ories, Observatoire de Paris, Universit\'{e} PSL, Universit\'{e} Paris Cit\'{e}, CNRS, F-92190 Meudon, France}

\date{Received 04/01/2022 Accepted 21/03/2022}

\abstract{
Neutron stars in low-mass binary systems are subject to accretion when material originating from the companion star accumulates on the surface. In most cases, the justified and common assumption in studying the properties of the neutron star crust is the fully accreted crust approximation. However, observations of some X-ray transient sources indicate that the original crust has not been completely replaced by accreted material, but is partly composed of the compressed original crust.
}
{
The crust of an accreting neutron star beyond the fully accreted crust approximation was studied; a two-part (or hybrid) crust made of the original crust that is compressed and of the accreted material crashing onto it was reconstructed as a function of the accretion stage. The differences in the composition and energy sources for the fully accreted and hybrid crusts influence the cooling and transport properties.
} 
{
A simple semi-empirical formula of a compressible liquid drop was used to compute the equation of state and composition of the hybrid crust. Calculations were based on the single-nucleus model,  with a more accurate treatment of the neutron drip point.
We compared the nuclear reactions triggered by compression in the original crust and in the accreted matter part of the hybrid crust.  We discuss another crust compression astrophysical phenomenon related to spinning neutron stars. 
}
{
The compression of the originally catalyzed outer crust triggers exothermic reactions (electron captures and pycnonuclear fusions) that deposit heat in the crust. 
The heat sources are cataloged as a function of the compression until the fully accreted crust approximation is reached. The pressure at which neutron drip occurs is a nonmonotonic function of the depth, leading to a temporary neutron drip anomaly. The additional potential source of energy for partially accreted crusts is the occurrence of a density inversion phenomenon between some compressed layers.
}
{
The original crust of a neutron star cannot be neglected for the initial period of accretion, when the original crust is not fully replaced by the accreted matter. The amount of heat associated with the compression of the original crust is on the same order of magnitude as that from the sources acting in  the accreted part of the hybrid crust.
} 


\titlerunning{Partially accreted crust}
\authorrunning{L.Suleiman}

\maketitle

\section{Introduction}
The composition and state of matter of the innermost layers of neutron stars (NS) remain a mystery that is actively researched by the nuclear astrophysics community. The outermost layers can be constrained by laboratory measurements of the mass of nuclei, ensuring that part of the crust is established with certainty. To explore deeper layers, nuclear experiments try to reproduce cold NS matter under conditions of density that are increasingly extreme; nevertheless, no experimental validation of the crust equation of state (EoS) beyond the neutron drip line, where free neutrons spill out of nuclei, has been established. Therefore, comparing observations of astrophysical parameters to their modeling (which directly depends on the EoS) can help us probe the nature of matter at high densities, and push the limits of nuclear laboratory experiments. 

Neutron star binary systems can host the process of accretion: matter from the companion star, whether it is a low-mass star or a high-mass star, is transferred to the NS surface. In the case of a low-mass X-ray binary system, the companion is either a white dwarf, a main-sequence star, or a star in the last stages of its life. The companion evolves into a donor when its Roche lobe has been filled and transfers matter to a disk orbiting the NS; a loss of angular momentum in the accretion disk then triggers the falling of this material onto the surface of the NS. In the case of a high-mass X-ray binary system, the companion is a massive star, and the mass transfer is either due to stellar wind or Roche-lobe overflow. The accretion process is a highly luminous phenomenon observed in X-rays with values of the exhibited luminosity up to 10$^{39}$erg/s for low-mass X-ray binaries. This process is intermittent, with short and luminous stages that can last from days to weeks, interspersed by quiescence stages (absence of accretion) that can last from months to decades. The sources are called soft X-ray transients, and their emission is observed in both accretion and quiescent stages, thus allowing astro-nuclear physicists to study the thermal evolution of NSs \citep[see][]{Wijnands2017}.

The temperature exhibited by NSs after accretion has stopped is higher than expected for isolated NSs, thus suggesting that accretion has induced heating processes deep enough below the surface to be still visible in quiescence. X-ray bursts are observed as bright flashes in low-mass X-ray binaries, with a rising time ${\sim 1~{\rm s}}$ and a typical duration of ${\sim 1~{\rm min}}$.  Observed X-ray bursts are powered by runaway burning of helium due to thermally unstable  $3\alpha$ fusion ignited at the bottom of the freshly accreted plasma. Thermonuclear burning, quenched by the exhaustion of helium and the decrease in temperature from the peak value exceeding $10^9$\;K, produces ashes of isotopes with $A=$\;50-110. The current understanding of the mechanism of X-ray bursts and their theoretical models is described in  \citet{Bildsten1998}, \citet{Parikh2013} and \citet{Meisel_2018}. The typical recurrence time for X-ray bursts is a few hours. As matter from an accretion disk is falling onto the surface, a series of exothermic reactions are triggered in the crust; heat sources are associated with these reactions, and this phenomenon is referred to as deep crustal heating \citep[see][]{Brown1998}. The generated heat is transported in the star and shows up partly as an additional surface luminosity detectable  in the band of soft X-rays. Observing the cooling of soft quiescent X-ray transients ($ \text{about ten}$ sources with observed thermal relaxation have been reported) is a chance to probe the nature of NS composition and EoS, as deep crustal heating directly depends on it.

The common approximation to establish the EoS and composition of accreted NSs is that of a fully accreted crust: the original crust, that is to say, the crust before accretion starts, is not considered because it is assumed to have been completely replaced by accreted material. The crust of a typical NS amounts to a total mass $M_{\rm crust} \sim 10^{-2}$ solar mass (M$_{\odot}$) \citep[see][]{Chamel2008}; the typical time of accretion $t_{\rm oc}$ to replace the original crust is  
\begin{equation}
    t_{\rm oc} = \frac{M_{\rm crust}}{\langle\dot{M}\rangle} \sim \frac{10^8}{\langle\dot{M}\rangle_{-10}} \hspace{0.2cm} \text{yr} , 
\end{equation}
where ${\langle \dot{M} \rangle_{-10}}$ is the time-averaged accretion rate in ${10^{-10}\;{\rm M}_{\odot}}$ per year (for values of $\dot{M}$ and ${\langle \dot{M} \rangle}$ for different sources, (see \citealt{Degenaar2015-2} and \citealt{Wijnands2017}). This accretion rate is defined as the conserved rest mass of diluted gas of baryons infalling onto the NS, per unit time, as measured by a distant observer. As binary systems involving NSs can exist  for as long as $10^9$ years, a fully accreted crust is a reasonable approach.

Even though the fully accreted crust approximation has been successful when some soft quiescent X-ray transients were modeled, the thermal relaxation of other sources suggest that this approximation cannot be applied \citep[see][]{Brown_2009}. One of these sources is IGR J17480$-$2446 \citep{Bonanno2015}, which has been observed before accretion, then during a two-month outburst in 2010, and two months after the accretion had stopped. It presents a low-frequency spin of 11 Hz \citep[see][]{Degenaar2015}, which suggests that the star's rotation has not been recycled by accretion via the angular momentum transfer of the orbiting accretion disk (see \citet{Recycled2020} and reference therein). If this source has only accreted a small amount of matter \citep{Degenaar2015-2}, the original crust has not been fully replaced, and the star would present a partially accreted crust. In this case, a hybrid crust made of the original crust as it is pushed toward the core, and of the above accreted material needs to be studied. The thermal relaxation of 1RXS J180408$-$342058 has been observed during its 4.5-month outburst in 2015 prior to quiescence (\citealt{Baglio2016} and \citealt{Marino2019}). \citet{Parikh2017} suggested that the luminosity of the source can be explained by a hybrid crust. The role of a hybrid crust in the thermal properties of accreting NSs has recently been studied without a detailed description of a compressed original, catalyzed crust \citep[see][]{Potekhin2021}.

Compression of the crust leading to exothermic reactions appears in astrophysical phenomena other than a low accretion rate in a binary. Recently, \citet{chamel2021} have proposed an analytical estimation, as well as numerical computation of heat sources and their location for magnetars with a decreasing magnetic field. The Lorentz force induced by a magnetic field balances the gravitational force, and its decrease is equivalent to a compression of the star. The slowing-down of a spinning NS as studied by \citet{IidaSato1997} can also lead to crust compression: a decrease in rotation rate leads to a change in the oval shape of the NS to a more spherical one. Because the crust of a star flattened by rotation is significantly thicker (i.e., contains more baryons) than that of a spherical star, spinning-down is equivalent to a relative compression in the shells of the crust.

In this paper, we follow the evolution of a nonaccreted (catalyzed) outer crust that is sinking toward the core as it is compressed under accreted material. In the second section, the method for calculating the EoS and composition of the catalyzed outer crust in a simple compressible liquid-drop nuclear model at zero temperature is described; it is also studied in detail  for accreted material using the same nuclear model. Reactions triggered by the accretion process in the partially accreted crust are presented. Results for the composition and EoS of the hybrid crust made of the catalyzed and accreted material are presented in Sect.~\ref{sec:results}. The appearance and disappearance of  heat sources as a function of the increase in pressure felt by the original outer crust are reconstructed for a partially accreted crust as well as for spinning-down compression. The properties of the originally catalyzed compressed outer crust are discussed in Sect.~\ref{sec:properties}. The timescales of compression-related phenomena discussed in the paper are estimated. We discuss the anomaly that occurs when the outer crust is compressed up to a pressure at which neutrons drip out of nuclei. Finally, we mention the density-inversion related instability triggered by nuclear reactions.

\section{Methods}
We focus on the evolution of an originally catalyzed outer crust under a uniform increase in pressure denoted $\Delta P$ (for partially accreted crust), and a relative increase in pressure denoted $\delta P$ (for a spinning-down NS). The study is restricted to a radial increase in pressure in spherical symmetry. At ${\Delta P=0}$, the crust is that of an isolated NS, and the EoS is calculated for cold catalyzed matter at global equilibrium, that is to say, in the ground state. The NS envelope is neglected in the sense that the EoS and composition are established from the mass density ${\rho \sim 10^6\;{\rm g/cm^3}}$; its importance in the cooling of soft quiescent X-ray transient sources is not forgotten, but the infalling accreted material has undergone enough transformations to be considered entirely made of $^{56}$Fe when it reaches the crust surface. The Wigner-Seitz cell approximation is used in the framework of the single nucleus model: Wigner-Seitz cells consist of a spherical and positively charged nucleus permeated  by a quasi-uniform  gas of relativistic electrons. In the inner crust, nuclei are additionally immersed in a neutron gas. In the following, $A_{\rm cell}$ refers to the number of nucleons per cell and $A$ to the number of nucleons in nucleus, $N$ is the number of neutrons inside nucleus, and $N_{\rm out}$ is the number of neutrons outside them.

\subsection{Catalyzed matter} \label{sec:catalyzedCrust}

We considered the outer crust of an isolated NS that has had time to cool after its proto-NS stage: temperature was neglected because matter is degenerate. A number of modern EoSs calculated at ground state exist\footnote{see the \href{https://compose.obspm.fr/}{CompOSE} data base \citep{Oertel2017}} based on microscopic many-body calculations or effective semi-phenomenological models such as the relativistic mean field or Skyrme forces theories. We chose a compressible liquid-drop model of nuclei. The Gibbs energy per cell ${G_{\rm cell}(P,A_{\rm cell},A, Z)}$, with $P$ the  pressure, follows the semi-phenomenological formula presented in \citet{Mackie1977}. At each pressure $P$, $G_{\rm cell}$ was evaluated for a large set of ${(A_{\rm cell}, A, Z)}$, and the combination of $A_{\rm cell}$, $A,$ and $Z$ with minimum Gibbs energy corresponds to the ground state; a region of the EoS in which  ${(A_{\rm cell},A, Z)}$ is stable is referred to as a $(A_{\rm cell},A, Z)$ shell. Recent laboratory measurements of masses of nuclei have been collected in the 2016 Atomic Mass Evaluation table (AME2016) \citep{ame2016}, thus gathering information on more than 2000 nuclei. If a combination ${(A_{\rm cell},A,Z)}$ is included in the AME2016, we replaced the calculated binding energy by the laboratory measurement. Evaluation of $G_{\rm cell}$ includes free neutrons if found outside nuclei, which determines the neutron drip point $P_{\rm ND}$, that is to say, the limit of the outer crust.

\subsection{Originally catalyzed compressed outer crust}

\begin{figure}
\resizebox{\hsize}{!}{\includegraphics{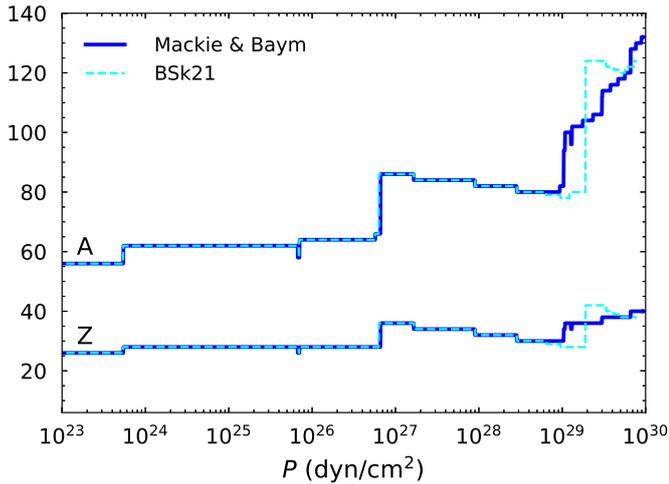}}
\caption{Catalyzed outer crust composition as a function of pressure $P$ in\;dyn/cm$^2$ as calculated in the Mackie \& Baym and BSk21 nuclear models.}
\label{fig:composition}
\end{figure}

To establish the composition and EoS of an originally catalyzed outer crust under compression, each shell at ground state as calculated in Sect.\ref{sec:catalyzedCrust} was taken out of global equilibrium to a local equilibrium: the Gibbs energy at a given pressure is constrained by its original catalyzed shell. Calculations are much faster because we need not estimate the Gibbs energy for all existing nuclei, but only the Gibbs energy of the original cell of the catalyzed shell $(A_{\rm cell}, A,Z)$, with respect to which the mentioned shell was subject to a compression related reaction.

\begin{figure*}
\resizebox{\hsize}{!}{\includegraphics{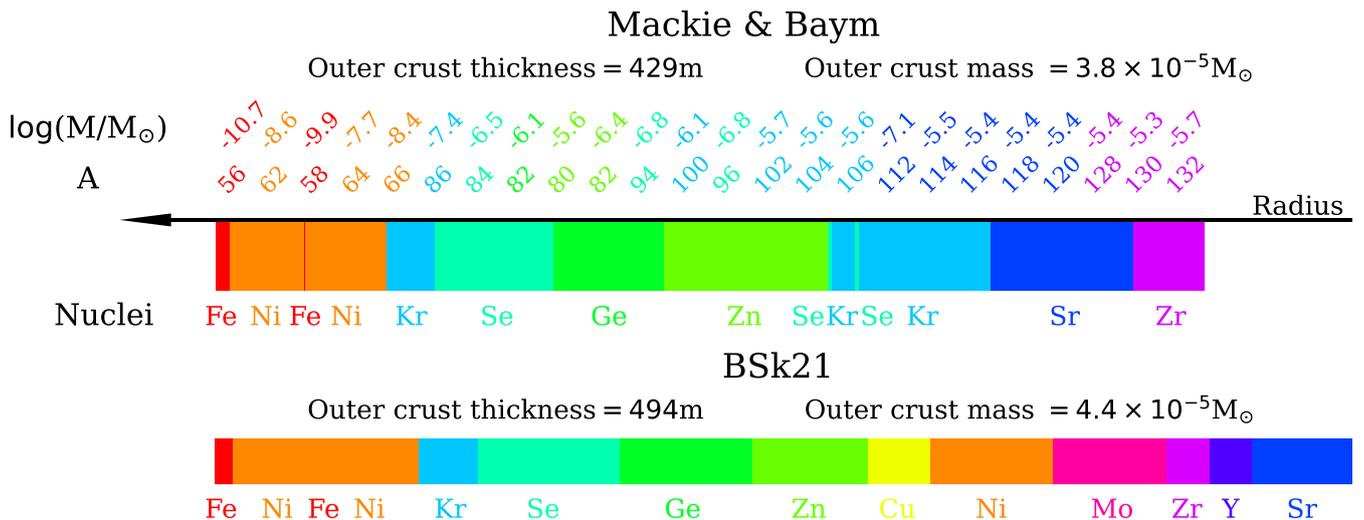}}
\caption{Structure of Mackie \& Baym outer crust compared to that of BSk21. Colors divide areas with different proton numbers (nuclei). For Mackie \& Baym, the nucleon number A is presented to give a complete definition of the shell $(A_{\rm cell},A,Z)$, for which the decimal logarithm of the gravitational mass in solar mass is presented. The mass and radius of the shells are computed using Eq.~\ref{eq:tovcrustm} and Eq.~\ref{eq:tovcrustr} for a total 1.4\;M$_{\odot}$ of the star, which corresponds to a radius for the Mackie \& Baym and BSk21 total radius of 11.7\;km and 12.6\;km, respectively.}
\label{fig:catstructure}
\end{figure*}

The first and most common compression-related reactions on nuclei (A,Z) are electron captures,  
\begin{equation}\label{eq:ec}
    (A,Z) + e^- \to (A,Z-1) + \nu_e,
\end{equation}
with electron $e^-$ and neutrino $\nu_e$. Because of the pairing term in the Mackie \& Baym model, a second electron capture occurs right after the first,
\begin{equation}\label{eq:ec2}
    (A,Z-1) + e^- \to (A,Z-2) + \nu_e + Q_{\rm ec}.
\end{equation}
The second electron capture is significantly exothermic, with released heat $Q_{\rm ec}$ up to $\sim 60$ keV per nucleon. Above the neutron drip line, these reactions can be accompanied by neutron emission, 
\begin{align} \label{eq:ec3}
    &(A,Z) + e^{-} \to (A-N_{1,\rm out}, Z-1) + \nu_{e} + N_{1,\rm out} \\
     &(A-N_{1,\rm out},Z-1) + e^- \to \nonumber\\
     &~~~~(A -N_{1,\rm out} -N_{2,\rm out},Z-2)+ \nu_e + Q_{\rm ec} + N_{2, \rm out},
\end{align}
where $N_{i, \rm out}$ is the number of emitted neutrons. 

The second type of compression-related exothermic reaction that can occur is pycnonuclear fusion: when electron captures have enriched the crust with neutrons, concurrently decreasing the number of protons per cell, the Coulomb repulsion fixing nuclei on the crust lattice is challenged by the high density, and fusion of two nuclei is favored,
\begin{equation}\label{eq:pycno}
    (A,Z) + (A,Z) \to (2A,2Z) + Q_{\rm pyc}; 
\end{equation}
this reaction can also be accompanied by neutron emission. The first pycnonuclear fusion occurs over the neutron drip line for $Z=10$. It should be stressed that the rate of pycnonuclear reactions is subject to very large uncertainty \citep{Yakovlev2006}. As a result, the nucleus for which pycnonuclear fusion occurs and the pressure (and density) at which this energy source is located are poorly defined. However, some global quantities such as the total energy release do not depend significantly on this rate \citep[see][]{Haensel2008}. We used the approach presented in \citet{Haensel2003} with the estimate of the pycnonuclear reaction timescale given by \citet{Sato1979}. Above the neutron drip line, but before pycnonuclear fusion sets in,  $A_{\rm cell}$ stays constant, but $A$ does not.

We denote by $P_{\rm bot}^{\rm cat}$ and $P_{\rm top}^{\rm cat}$  the bottom (highest) and top (lowest) pressure boundaries of the catalyzed shell $(A_{\rm cell},A,Z)$, and $\Delta P$ in dyn/cm$^2$ is the increase in pressure (compression) felt by the original crust. After a given compression $\Delta P$, the above-mentioned shell has evolved to pressure boundaries ${P_{\rm bot}^{\rm cat} + \Delta P}$ and ${P_{\rm top}^{\rm cat} + \Delta P}$. As exothermic reactions occur in the shell under compression, the outer crust has evolved from that catalyzed state to a local equilibrium state at given $\Delta P$; in this sense, a compressed crust is a reservoir of heat at local equilibrium.

The mass of a typical catalyzed crust is ${M_{\rm c} = 10^{-2}\;{\rm M}_{\odot}}$ \citep[see][Sect. 4.3]{Chamel2008}. In the framework of partially accreted crusts, the increase in pressure needed to replace the crust by accreted material and reach the fully accreted crust approximation is ${\Delta P \sim 10^{32}\;{\rm dyn/cm^2}}$. Our calculations are given up to that value. At ${\Delta P \sim 10^{32}\;{\rm dyn/cm^2}}$, the outer crust has been pushed to the bottom of the inner crust, close to the core, reaching a density of ${\sim 5 \times 10^{13}\;{\rm g/cm^3}}$.

\subsection{Accreted material part of the partially accreted crust}

Partially accreted crusts  are made of the original crust that is compressed and of the accreted material falling on its top and supplying the compression. The hydrogen- and/or helium-rich plasma originating from the companion star is falling onto the NS, forming an envelope. The kinetic energy of the plasma ions transformed into heat powers the quiescent radiation between the bursts. The freshly accreted hydrogen- and helium-rich plasma is compressed under the weight of newly accreted  matter from the companion star supplied by the accretion disk. Accreted matter accumulates in a layer of growing mass, typically during a few hours, and the pressure at its bottom increases. The subsequent evolution of the accreted layer depends on the accretion rate  ${\dot{M}={\dot{M}}_{-10} 10^{-10}~{\rm M}_{\odot}~{\rm yr^{-1}}}$ and metallicity of freshly accreted material \citep{Bildsten1998}; more recent reviews of nucleosynthesis in X-ray bursts can be found in \citet{Parikh2013} and \citet{Meisel_2018}. For  ${2<{\dot{M}}_{-10}<10}$, hydrogen burns stably into helium. When no hydrogen is left in the accreted bottom layer, the pure helium layer starts to grow. Helium burning, through the $3\alpha$ fusion, initiated after crossing the ignition line in the density-temperature plane, is associated with thermal instability, resulting in thermonuclear runaway. The accreted envelope is then heated, with a peak temperature $ \sim 10^{9}~{\rm K}$ reached in a second. For ${10<{\dot{M}}_{-10}<260}$, helium is ignited before hydrogen burning has been completed, such that thermal instability and thermonuclear runaway of helium via $3\alpha$ fusion takes place in the hydrogen-helium mixture. In both accretion regimes, nucleosynthesis is driven mainly by the rapid proton captures and following $\beta^+$ decays, but other nuclear processes involving $\alpha$ and protons are also involved. Generally, nuclear ash abundances peak around $A\sim 60$ and/or $A\sim 100$, depending on the accretion rate and metallicity of the accreted material \citep{Parikh2013}.

The sequence (helium flash, then the formation of the thermonuclear ash layer) is repeated during accretion episodes in soft X-ray transients (with a typical recurrence time of a few hours). As the process continues, ashes accumulate and are pushed to higher pressures by freshly accreted material. We approximate thermonuclear ashes by pure $^{56}$Fe. Thermonuclear ashes approximated by $^{106}$Pd  were considered in \citet{Haensel2008}.

The same scheme as in the previous section was used to calculate the accreted material EoS: stability of ${(A_{\rm cell}, A,Z)}$ nuclei with respect to electron capture, and pycnonuclear fusion was evaluated under an increasing pressure. This approach was taken in the fully accreted crust approximation, and we designate this area of the partially accreted crust as the accreted material part of the crust. 

The hybrid crust was reconstructed at each stage of the compression characterized by $\Delta P$ by adding the accreted material part at the top of the first shell of the originally catalyzed compressed outer crust. The heat sources associated with exothermic reactions in the accreted material part are permanent: they remain active at a fixed pressure in the partially accreted crust because accreted material falls down and is compressed continually. Exothermic reactions in the originally catalyzed compressed outer crust appear and disappear, depositing heat in the matter for as long as the shell goes from its original state of ${(A_{\rm cell}, A,Z)}$ to the daughter nuclei ${(A_{\rm cell}', A', Z')}$. For a comprehensive study of partially accreted crusts, we created a dynamic catalog of heat sources.

\subsection{Neutron drip}

Above a certain density, nuclei are subject to neutron emission: this is the neutron drip phenomenon. In the framework of the single-nucleus model, the neutron drip point is given by the solution to the equation
\begin{equation}
    G_{\rm cell}(A_{\rm cell},A,Z) = G_{\rm cell}(A_{\rm cell}, A -N_{\rm out}, Z-1),
\end{equation}
with $N_{\rm out}$ the number of neutrons outside the nucleus. We note that an electron capture is evaluated just before the neutron emission: the timescale of strong interaction is much shorter than that of weak interaction. The first fulfills the energy requirement imposed by the weak process, therefore neutron emission is always triggered by an electron capture.

The single-nucleus model is equivalent to the assumption that $N_{\rm out}$ neutrons drip simultaneously in all Wigner-Seitz cells at given pressure $P_{\rm nd}$. As a consequence, the number density of the neutron gas outside nuclei $n_{\rm out}$ is discontinuous at $P_{\rm nd}$. For ${P>P_{\rm nd}}$, $n_{\rm out}$ is larger by a factor $N_{\rm out}$ than the number density of nuclei ${n_{\cal N} = 1 / V_{\rm cell}}$, with $V_{\rm cell}$ the volume of the cell. To accurately determine the neutron drip point, we need to go beyond the single-nucleus model as presented in \citet{Chamel2015}, referred to below as the continuous approach: the density of the neutron gas is assumed to be negligible at the neutron drip point, and the only energy considered for free neutrons is their rest energy. The chain of reactions corresponds to
\begin{align}\label{eq:ndonset}
    &(A_{\rm cell},A,Z) \to (A_{\rm cell}-N_{\rm out},A-N_{\rm out}, Z-1) + N_{\rm out}, 
\end{align}
for which the number of nucleons in the Wigner-Seitz cell decreases by $N_{\rm out}$. The onset of neutron drip occurs when the Gibbs energy of left and right nucleus in Eq.~\eqref{eq:ndonset} follows
\begin{align}
    &G_{\rm cell}(A_{\rm cell},A,Z) = \nonumber \\
    &G_{\rm cell}(A_{\rm cell}-N_{\rm out},A- N_{\rm out},Z-1,) + m_nc^2 {N_{\rm out}.}
\end{align}

In our model, we considered only ground-state to ground-state transitions driven by electron captures and associated with neutron emission. Electron captures leading to the excited state of the daughter nucleus that de-excite by neutron emission were considered in \citet{Gupta2008} and \citet{Lau2018}. These effects are not taken into account in this paper. Neutron emission driven by the photon absorption by nuclei is much less important because at temperature ${T\simeq 10^{8}\;{\rm K}}$ and ${\rho \simeq 10^{11}\;{\rm g/cm^3}}$, plasmon suppression of photons is too strong \citep{Gupta2008}.

\section{Results}\label{sec:results}

\begin{figure}
\resizebox{\hsize}{!}{\includegraphics{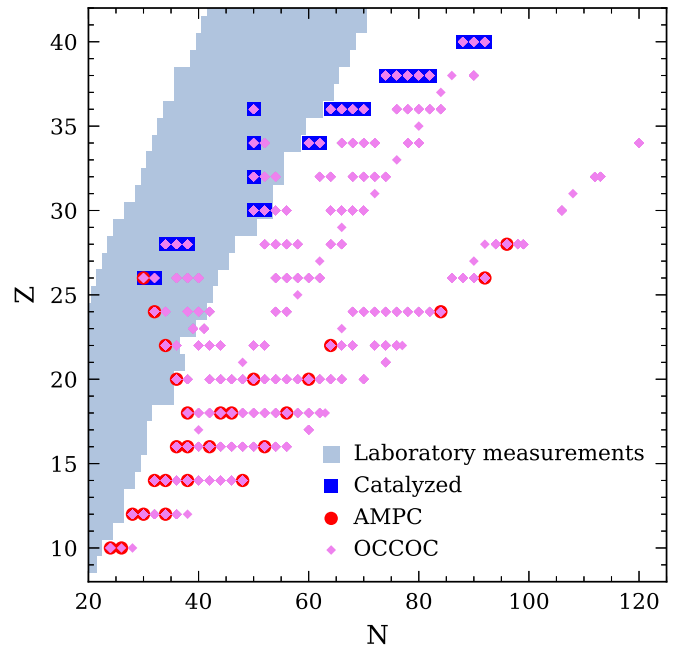}}
\caption{Abundance of nuclei $(N,Z)$ found in the partially accreted crust calculated using the  Mackie \& Baym model, compared to the valley of recently measured nuclei from AME2016. $(N,Z)$ for the catalyzed outer crust are shown in blue, $(N,Z)$ of the accreted material part of the crust (AMPC) are shown in red, and $(N,Z)$ for the originally catalyzed compressed outer crust (OCCOC) are shown in violet.}
\label{fig:abundance}
\end{figure}

\subsection{Catalyzed outer crust and its compression}
\label{sect:resoc}

The composition of the catalyzed outer crust calculated with the Mackie \& Baym model is presented in Fig.~\ref{fig:composition}. It is made of 24 shells, with a neutron drip line at ${P_{\rm nd} \simeq 1.1 \times 10^{30}\;{\rm dyn/cm^2}}$, and a last shell of $^{132}$Zr. In this figure, it is compared to the outer crust calculated in the nonrelativistic phenomenological approach of Brussels-Skyrme established by \citet{Pearson2018}. Brussels-Skyrme 21 is a functional density model parameterized to fit laboratory measurements of nuclei: it includes considerations of the proton shell effects as well as neutron skin effects. Its catalyzed outer crust consists of 18 shells, some with an odd proton number, and a neutron drip line at ${P_{\rm nd} \simeq 7.8\times 10^{29}\;{\rm dyn/cm^2}}$. The models of Mackie \& Baym and Brussels-Skyrme 21 are similar up to ${P \simeq 6\times 10^{28}\;{\rm dyn/cm^2}}$, with $^{80}$Zn the last common nucleus; up to this pressure, compositions are calibrated to experimentally determined masses of nuclei AME2016. For the next shell, the Gibbs energy of the odd proton number element $^{79}$Cu calculated by Brussels-Skyrme 21 is lower than the calibrated $^{82}$Zn that was selected in Mackie \& Baym; from then on to deeper shells, the compositions differ. One thin shell of $^{58}$Fe appears for Mackie \& Baym and Brussels-Skyrme 21: as discussed in Sect.~7.4.1 of \citet{Blaschke2018}, this shell does not exist when a particular consideration for electron charge polarization is taken.

\begin{figure}
\resizebox{\hsize}{!}{\includegraphics{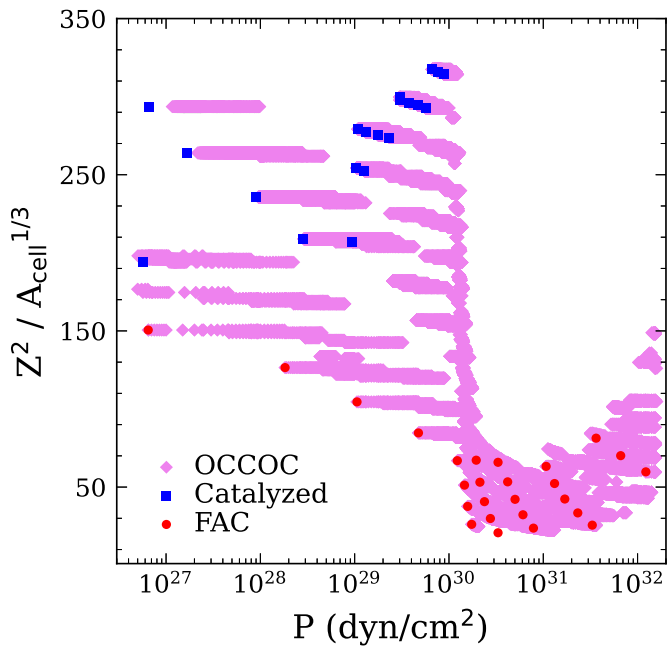}}
\caption{$Z^2/A^{1/3}$ as a function of the pressure $P$ in dyn/cm$^2$ for the catalyzed outer crust is shown in blue, for the fully accreted crust (FAC), it is shown in red, and for the originally catalyzed compressed outer crust (OCCOC), it is shown in violet. $Z$ is the proton number, and $A$ is the nucleon number per nucleus. }
\label{fig:ZoverA}
\end{figure}

The mass and thickness of each shell are presented for the Mackie \& Baym and Brussels-Skyrme 21 models in Fig.~\ref{fig:catstructure}. The thickness and the gravitational mass of the shells for a given stellar configuration can be approximated with high accuracy from formulas established in \citet{Zdunik2016},
\begin{equation} \label{eq:tovcrustr}
  \Delta R = 73\,{\rm m}\; \Delta \mu_{\rm MeV}\;\frac{R^2_{6}}{M/M_\odot} \bigg(1-0.295\frac{M/M_\odot}{R_{6}}\bigg) ,\\
\end{equation}
\begin{equation} \label{eq:tovcrustm}
    \frac{\Delta M}{M_\odot} = 0.47\cdot 10^{-4}\;\Delta P_{30}\frac{R^4_{6}}{M/M_\odot} \bigg(1-0.295\frac{M/M_\odot}{R_{6}} \bigg),
\end{equation}
where $R_6$ is the NS radius divided by 10 km, $\Delta \mu$ and  $\Delta P$ correspond to the thickness of the shell (crust) in baryon chemical potential and pressure, respectively; those quantities are given in MeV and ${10^{30}\;{\rm dyn/cm^2}}$ respectively. In Fig.~\ref{fig:catstructure}, a total NS mass of ${1.4\;{\rm M}_{\odot}}$ was selected for the two nuclear models of dense matter SLy4 with the Mackie \& Baym crust and Brussels-Skyrme 21 (unified EoS for core and crust), to which corresponds a total radius of the star of 11.7\;km  and 12.6\;km,  respectively. The outer crust of the SLy4/Mackie \& Baym model is thinner than that of the Brussels-Skyrme 21 model, 429 m and 494 m, respectively. The ratio of the outer crust thickness is equal to the square of the ratio of the respective stellar radii, see Eq.~\ref{eq:tovcrustr}. The accreted crust for the Mackie \& Baym model is made of 36 shells, including 5 in the outer crust, for a total thickness of 722 m and a mass of ${3\times 10^{-3}\;{\rm M}_{\odot}}$.

The set of nuclei found in the originally catalyzed compressed outer crust is displayed in Fig.~\ref{fig:abundance}, with that of the accreted material and that of the catalyzed outer crust. There are around 170 nuclei appearing in the originally catalyzed compressed outer crust; they overlap with the catalyzed nuclei,\ as catalyzed matter is the stage of the originally catalyzed compressed outer crust at $\Delta P=0$, and also with the accreted material nuclei when the stage of compression reaches the fully accreted crust approximation.

The composition of the originally catalyzed compressed outer crust differs from that of the fully accreted crust and from that of the catalyzed one. The composition of the partially accreted crust is displayed in the lower figure of the movie (\textit{movHeat.mpg}) presented in Ancillary files; the composition of the catalyzed outer crust is also presented as a landmark. The accreted material part is seen to evolve from low to high density (left to right), thus invading the partially accreted crust until it dominates at ${\Delta P=10^{32}\;{\rm dyn/cm^2}}$. Thermal and transport properties of the crust depend on the crust composition. For example, the melting temperature and shear modulus are both proportional to ${Z^2/A_{\rm cell}^{1/3}}$ \citep[see][]{Chamel2008}; this quantity is presented as a function of the pressure in Fig.~\ref{fig:ZoverA}. The melting temperature in the fully accreted crust is much lower than that of the originally catalyzed compressed outer crust: for a range of temperatures, the fully accreted crust is liquid and the originally catalyzed compressed outer crust is solid. The shear modulus is also larger for the originally catalyzed compressed outer crust than for the fully accreted crust: the former is more rigid than the latter.

\begin{figure}
\resizebox{\hsize}{!}{\includegraphics{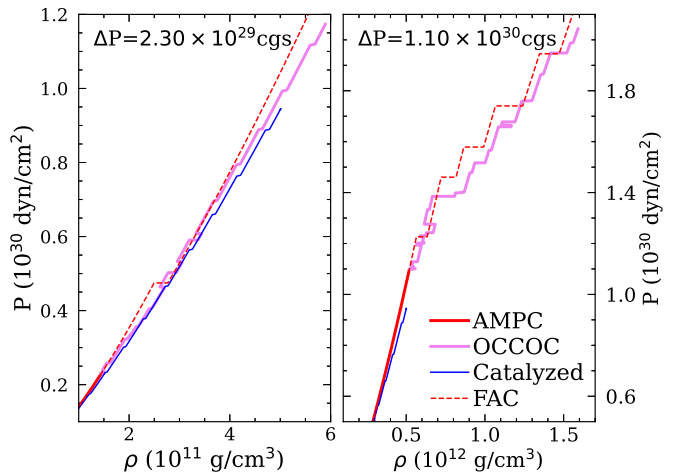}}
\caption{Equation of state P($\rho$) of the partially accreted crust for compression ${\Delta P=2.3\times 10^{29}\;{\rm dyn/cm^2}}$ and ${\Delta P =1.1 \times 10^{30}\;{\rm dyn/cm^2}}$. The originally catalyzed compressed outer crust (OCCOC) is presented in violet, and the accreted material part of the crust (AMPC) is presented in red. The catalyzed outer crust EoS is presented in blue, and the fully accreted crust (FAC) EoS is shown as dashed red lines. For the density inversion in the EoS of the originally catalyzed compressed outer crust, see Sect.~\ref{sec:drinstablity}.}
\label{fig:eosPAC}
\end{figure}

The composition of the crust also affects the modeling of the thermal evolution of accreting NSs. On the one hand, the thermal conductivity of the crust is mostly contributed by the electrons; it is inversely proportional to the scattering frequency of electrons, which is proportional to $Z^2$. The proton number of the originally catalyzed compressed outer crust being generally larger than that of the fully accreted crust at fixed density, thermal conductivity is lower and the crust exhibits a lower heat transfer for the originally catalyzed outer crust than for the fully accreted crust. On the other hand, the heat capacity of the crust measures the energy required to raise the temperature by one Kelvin; it enters the relativistic balance equations. In the crust, the electron contribution scales as ${\sim Z^{2/3} /A^{2/9}}$, and the lattice as $A^{1/3}$ \citep[see][]{Fortin2010}. 

The relation between the pressure and the density for the partially accreted crust is presented in a movie (\textit{movPR.mpg}) in Ancillary files; in Fig.~\ref{fig:eosPAC}, $P(\rho)$ is presented at compression ${\Delta P = 2.3\times 10^{29}}$ and ${1.1\times 10^{30}\;{\rm dyn/cm^2}}$. The equation of state for catalyzed matter is the softest as it corresponds to a global minimum of energy at a given pressure. The stiffest EoS is obtained for the fully accreted crust. In this case, the local energy minimum is determined for a relatively small number of nucleons in the Wigner-Seitz cell. This favors nuclei with $A$ smaller than in the case of the catalyzed and also originally catalyzed compressed matter. As a result, matter in the fully accreted crust is farther away from global equilibrium (catalyzed matter) and originally catalyzed compressed matter. Some quantities depend directly on the equation of state of the crust. Superfluidity depends on the chemical potential and on the energy density. Neutrino emission processes such as electron-electron Bremsstrahlung, electron-nucleus Bremsstrahlung, and neutrino pair-Bremsstrahlung can occur in the crust and depend on the chemical potential and the density. We note that the neutrino emission contribution of the crust is negligible compared to that of the core \citep[see][]{Fortin2018}.

\subsection{Heat sources of the originally catalyzed compressed outer crust}

Reactions induced by the compression of a shell $(A_{\rm cell}, Z, N)$ are exothermic and deposit heat in the originally catalyzed compressed outer crust. The first electron capture occurs for an increase in pressure ${\Delta P \simeq 7.5 \times 10^{27}\;{\rm dyn/cm}^2}$; the largest energy release per nucleon due to electron capture is  $\sim 110\;{\rm keV/A}$ at ${\Delta P \simeq 4.6 \times 10^{28}\;{\rm dyn/cm^2}}$. Pycnonuclear fusions are far more rare than electron captures, with a maximum energy release of ${\sim 340\;{\rm keV/A}}$ for an increase in pressure ${\Delta P \simeq 1.9 \times 10^{30}\;{\rm dyn/cm}^2.}$ This is also the first pycnonuclear reaction. It occurs in the first shell of the originally catalyzed compressed outer crust for parent nuclei ${(A_{\rm cell}=56,Z=10,N=24)}$. The details of the exothermic reactions in the originally catalyzed compressed outer crust up to the neutron drip point are presented in Tables~\ref{tab:pacoc1}, \ref{tab:pacoc2}, and \ref{tab:pacoc3} of the appendix.

The problem of the interplay of neutrino losses and matter heating deserves additional discussion. The electron capture chain is initiated by the process in Eq.\eqref{eq:ec} after the chemical potential of electrons $\mu_e$ reaches the threshold $\mu_e=W_1$, with $W$ the threshold of the reaction; this process proceeds in a quasi-equilibrium way (no heating). The nucleus $(A,Z-1)$ has odd numbers $N$ and $Z$ and therefore has a significantly lower threshold for the second electron capture: $W_2<W_1$. Moreover, an odd-odd nucleus $(A,Z-1)$ has a dense spectrum of excited states with excitation energies (relative to the ground state) denoted $E_{\rm exc}$, which follows $0<E_{\rm exc}<E_e-W_2$. An excited state $(A,Z-1)^*$ decays to the ground state $(A,Z-1)_{\rm gs}$. The important effect connected with the transition to the exited state of $(A,Z-1)$ nucleus is the increase in threshold density (pressure) due to the excitation energy $E_{\rm exc}$. This also leads to a larger energy release in the second reaction $(A,Z-1)\to (A,Z-2)$ and a maximum energy release increase that arises because the transition to the excited state is larger than the ground state - ground state  transition by $2E_{\rm exc}$ \citep[see][]{chamel2021}. The net contribution of excited states is found to more than balance the neutrino losses \citep[see][]{Gupta2007}. Therefore, the neutrino losses can be neglected, and our approximation ${Q_2\simeq \mu_e-W_2}$ is valid.

\begin{figure}
\resizebox{\hsize}{!}{\includegraphics{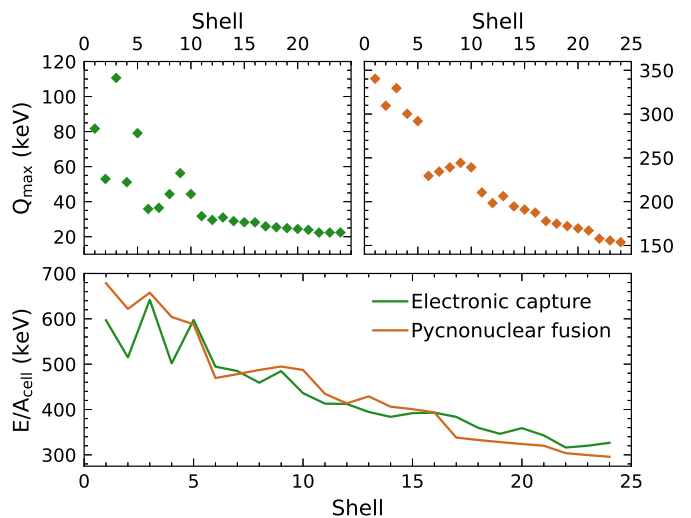}}
\caption{Maximum energy release $Q_{\rm max}$ per nucleon per shell for electron captures (upper left plot) and pycnonuclear fusions (upper right plot), as well as the sum of all electron captures and pycnonuclear fusion heat release per shell $E/A_{\rm cell}$ (lower plot). The shell numbers go from 1 at the outer crust surface to 24 at the outer or inner crust transition.}
\label{fig:heat_max}
\end{figure}

The maximum heat release per nucleon $Q_{\rm max}$ in each of the 24 shells during the compression of the originally catalyzed crust up to ${\Delta P = 10^{32}\;{\rm dyn/cm^2}}$ is presented in Fig.~\ref{fig:heat_max}. Generally, $Q_{\rm max}$ decreases when considering shells initially located deeper in the original crust. In the same figure, the total energy per nucleon $E/A_{\rm cell}$ released in a shell follows the same trend. The energy per nucleon is only defined per one shell of the compressed outer crust and not for the whole compressed outer crust: the energy per nucleon of different shells cannot be summed. The decrease in energy release per nucleon as a function of the depth in the star can be explained from the catalyzed composition of the outer crust, as displayed in Fig.~\ref{fig:composition}. The nucleon number $A$ is generally larger in deeper shells of the crust. Large nuclei are more stable than small ones; applying a compression to the shells of the catalyzed outer crust induces a state of local equilibrium that is closer to global equilibrium for shells with high nucleon number. In other words, catalyzed shells of the outer crust with large nucleon number are more stable with regard to compression-related reactions than shells with small nucleon number. Shells deeper in the crust release less energy per nucleon than shells close to the surface. A few shells do not follow this decrease, and this can be explained with the same reasoning: in Fig.~\ref{fig:heat_max}, the third shell (very thin layer of $^{58}$Fe;  its existence is uncertain, see Sect.~\ref{sect:resoc}) presents a higher maximum energy per nucleon than its shallower neighbor $^{62}$Ni; there is also a slight increase in $Q_{\rm max}$ for shells 6 ($^{86}$Kr), 7 ($^{84}$Se), 8 ($^{82}$Se), and 9 ($^{80}$Ge). These shells follow the catalyzed nucleon number logic stated above.

To calculate the total energy released in the originally catalyzed compressed outer crust, we allocated the exothermic reactions to the number of baryons that go through these reactions. The number of baryons was calculated using the Tolmann-Oppenheimer-Volkoff equations applied to the crust ${(P/\rho c^2 \ll 1}$ and ${PR^3/M c^2 \ll 1}$), as well as the expression for the baryon number in a thin spherical shell, ${\rm d}N_{\rm b}$ , of the coordinate thickness ${\rm d}r$. In General Relativity,

\begin{equation}
    {\rm d}N_{\rm b} = \frac{4 \pi r^2 n_B}{\sqrt{1-\frac{2Gm}{rc^2}}} {\rm d}r.
\end{equation}

Therefore, an  approximate number of baryons $\Delta N_i$ in an $i$-th shell of pressure range ${P^{i}_{\rm bot} - P^{i}_{\rm top}}$ is\begin{equation}
    \Delta N_i = \frac{4 \pi R^4}{GMm_0} \sqrt{1-\frac{2GM}{Rc^2}} \big( P^i_{\rm bot} - P^i_{\rm top} \big), 
\label{eq:deltaN}
\end{equation}
where ${m_0 = 931\;{\rm MeV/c^2}}$ is the mean bound nucleon mass in $^{56}$Fe. In contrast to the decreased energy per nucleon, the number of nucleons per shell generally increases deeper in the outer crust (see the gravitational mass of the shells in Fig.~\ref{fig:composition}).

The formula for the total energy released in the compressed outer crust reads\begin{align}
    E=&\sum_i \Delta N_i (\Sigma_j E_{ij})\nonumber\\ 
    =&\frac{4 \pi R^4}{GMm_0} \sqrt{1-\frac{2GM}{Rc^2}}\sum_i \Delta P_i(\Sigma_j E_{ij}), 
\label{eq:Eoc}
\end{align}
where $E_{ij}$ is j-$th$ energy source located in the i-$th$ shell, and ${\Delta P_i=P^i_{\rm bot} - P^i_{\rm top}}$ is the thickness in pressure of the i-$th$ shell. For the accreted material part of the crust, formula \eqref{eq:Eoc} simplifies to
\begin{equation}
    E=\frac{4 \pi R^4}{GMm_0} \sqrt{1-\frac{2GM}{Rc^2}}\sum_j P_j E_j .
\label{eq:Eocacc}
\end{equation}

We introduce the factor $\alpha^*$ , which depends on the NS mass and radius, 
\begin{align}
    \alpha_* &= \frac{R_{6}^4}{M/{M_\odot}} \sqrt{1 - 0.295\frac{M/{M_\odot}}{R_{6}}},
\end{align}
such that 
\begin{align}
    &\frac{4 \pi R^4}{GMm_0} \sqrt{1-\frac{2GM}{Rc^2}}= 9.3 \times 10^{16} \alpha^*\,{\rm\frac{cm^3}{MeV}} .
\end{align} 
For energy considerations, we assumed a $1.4\;{\rm M}_\odot$ NS, which corresponds to a radius of $11.7$\;km in the Mackie \& Baym nuclear model.

\begin{figure}
\resizebox{\hsize}{!}{\includegraphics{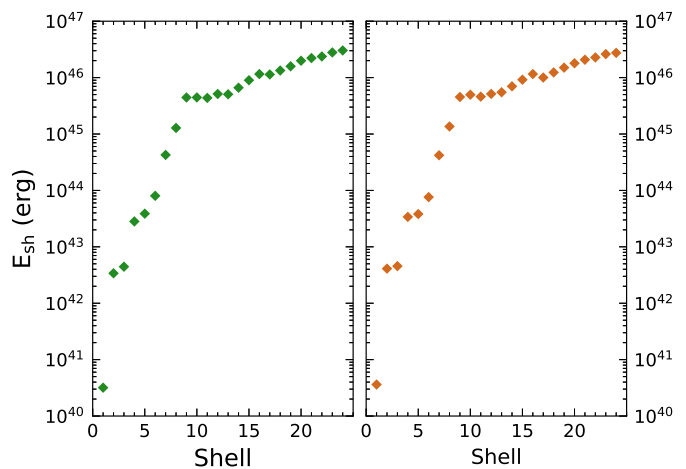}}
\caption{Total energy in erg released in each of the 24 shells of the originally catalyzed compressed outer crust for electron captures (left) and pycnonuclear fusions (right).}
\label{fig:ergShell}
\end{figure}

In Fig.~\ref{fig:ergShell} we present the total energy $E_{\rm sh}$ released in each of the 24 shells of the originally catalyzed compressed outer crust. The trend of a decrease in maximum energy per nucleon shown in Fig.~\ref{fig:heat_max} is largely compensated for by the number of nucleons in the shells, thus ensuring that most of the energy is released in the deepest shells of the compressed outer crust. Shell 17 ($^{112}$Sr) has a particularly small pressure range, which explains why it presents a low point of total energy in Fig.~\ref{fig:ergShell} and a low mass in Fig.~\ref{fig:composition}. The total energy released by the originally catalyzed compressed outer crust, up to ${\Delta P = 10^{32}\;{\rm dyn/cm^2}}$, corresponds to the sum of sources displayed in Fig.~\ref{fig:ergShell} for a total of ${{\rm E}_{\rm tot} = 4.25 \times 10^{47}\;{\rm erg}}$. 

Our calculations were made for an originally catalyzed outer crust. To compare it to that of an originally accreted outer crust, we calculated the total energy needed to sink the 24 shells of the originally catalyzed outer crust over their neutron drip point (Eq.~\eqref{eq:Eoc}) and the 5 shells of an originally accreted outer crust over their neutron drip point (Eq.~\eqref{eq:Eocacc}).
The total energy is ${7.8\times 10^{45}\;{\rm erg}}$ and ${7.4 \times 10^{45}\;{\rm erg}}$. These numbers indicate that the total energy available from the compression of the catalyzed and accreted crust is similar. In the case of the accreted crust, the energy released is continuous and the sources are located at fixed densities (five energy sources). For the originally catalyzed compressed outer crust,  many temporary sources exist (about two to five sources per each shell before neutron drip; see Fig.~\ref{fig:lasting} and Table~\ref{tab:pacoc1}).

To compress the original outer crust (originally catalyzed or accreted) into the inner crust, an accretion of the amount of matter equal to the outer crust mass is needed. In this case, the total energy released by the accreted material part of the crust is ${2.2\times 10^{46}\;{\rm erg}}$. This complete replacement of the initially catalyzed outer crust, now built of accreted matter, results in the pressure at the bottom of the original crust ${\sim 1.5 \times 10^{30}\;{\rm dyn/cm^2}}$. In addition to the ${7.8 \times 10^{45}\;{\rm erg}}$ released in the initial outer crust, this compression also leads to chains of reactions over the neutron drip point: the additionally released energy is $\sim 5 \times 10^{45}$\;erg. In total, the replacement of the outer crust by the accreted matter gives ${1.3\times 10^{46}\;{\rm erg}}$ in the originally catalyzed compressed outer crust.

\subsection{Onset of neutron drip}
\begin{figure}
\resizebox{\hsize}{!}{\includegraphics{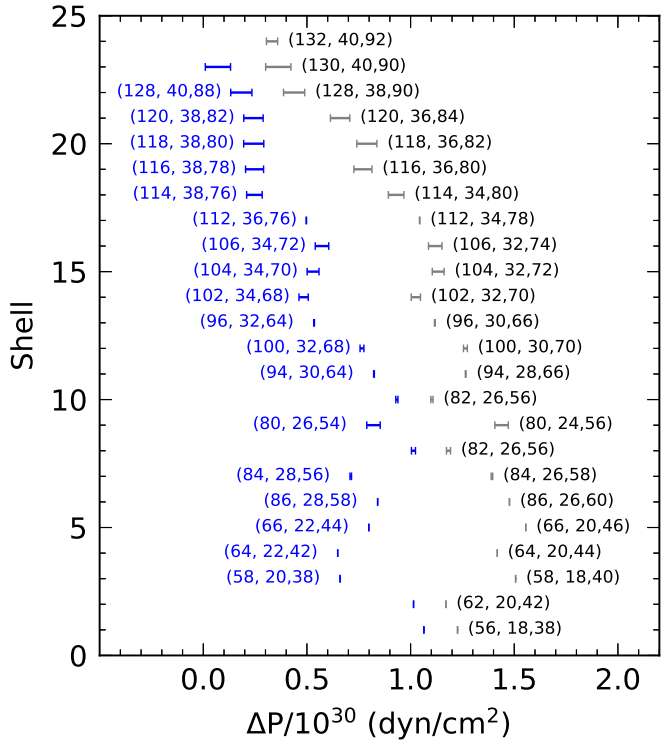}}
\caption{Compression required to reach the neutron drip point in the single-nucleus model (black) and the continuous approach (blue) for each of the shells of the originally catalyzed compressed outer crust. The error bars refer to the additional compression needed for the entire shell to undergo neutron emission from bottom (left end of the error bar) to top (right end of the error bar). The parent nuclei undergoing electron capture that precedes the neutron emission are indicated with the notation ${(A_{\rm cell}, Z,N)}$. In the case of a continuous approach, it is presented only if the parent nucleus is different than that in the single nucleus models.} 
\label{fig:NDcompression}
\end{figure}

The neutron drip point depends on the density and on the shell ${(A_{\rm cell},Z,N)}$. Because the catalyzed outer crust is made of 24 shells, in the originally catalyzed compressed outer crust, this point is different for each shell. The compression at which neutrons start dripping out of nuclei as well as the additional compression needed to produce free neutrons throughout the entire shell is displayed in Fig.~\ref{fig:NDcompression} for the 24 shells of the originally catalyzed compressed outer crust; this is also presented in Tables~\ref{tab:pacoc1}, \ref{tab:pacoc2}, and \ref{tab:pacoc3}. The exothermic reactions up to the neutron drip point calculated in the single-nucleus model are shown in  black, and those calculated with the continuous approach are shown in blue.

As an example, we considered the 19th shell of the originally catalyzed compressed outer crust, which corresponds to an original (catalyzed) shell of $^{116}_{38} {\rm Sr}$. In the single-nucleus model, the neutron drip begins at the top of the shell after a compression ${\Delta P \simeq 7.3 \times 10^{29}\;{\rm dyn/cm^2}}$, and requires an additional compression of ${8.7\times 10^{28}\;{\rm dyn/cm^2}}$ for the entire shell to pass from $^{116}$Kr to $^{106}$Ge. The mechanism consists of four electron captures triggering the neutron drip at a depth in pressure of ${\sim 1.2 \times 10^{30}\;{\rm dyn/cm^2}}$. The energy release associated with this process is about 35\;keV (from ${940.471 \to 940.435\;{\rm MeV}}$). The number density of neutrons $n_{\rm out}$ , trapped in one Wigner-Seitz cell at the onset, is an order of magnitude larger than that of nuclei and about only one order of magnitude smaller than the average baryon number density. In the continuous approach, the neutron drip begins earlier, at a compression ${\Delta P \simeq 2 \times 10^{29}\;{\rm dyn/cm^2}}$, with a double electron capture in the reaction $^{116}$Sr to $^{114}$Kr at a depth in pressure of ${\sim 6.7 \times 10^{29}\;{\rm dyn/cm^2}}$. The energy release associated with the reaction is about 13\;keV. By construction, in this approach, the average number density of free neutrons is negligibly small. 

Because the onset of the neutron drip occurs for an earlier (lower) compression in the continuous approach than in the single-nucleus model, the number of reactions in the outer crust is reduced: in the third shell, the neutron drip for the continuous approach occurs for the parent nuclei of the single-nucleus model. In shell 18, it occurs for the grandparent nucleus (see Tables~\ref{tab:pacoc1}, \ref{tab:pacoc2}, and \ref{tab:pacoc3}). For the same reason, the last shell of the outer crust for which the neutron drip is calculated in the continuous approach is shell 23 ($^{130}$Zr) of the outer crust, calculated within the single-nucleus approach.

\subsection{Spinning-down of the neutron star and the induced crust compression}

One of the scenarios leading to an increase in pressure in the NS crust is the process of slowing-down of the star's rotation. This case has been studied by \citet{IidaSato1997} and \citet{Gusakov2015}, and their result gives an increase in pressure by some $25\%$ for an initial frequency of 1 kHz. A similar conclusion can be drawn from the consideration  of the (baryon) mass of the outer crust for a rotating strange quark star \citep{Zdunik2001}. In the latter case, the relative increase in pressure is approximately equal to the relative increase in baryon mass of a given shell, which is about $30\%$ for an initial rotation at 1 kHz, and $\sim 100\%$ at most for a spinning-down from an initial rotating configuration close to a Keplerian one. For such a  relative increase in pressure, detailed calculations show the existence of exothermic reactions in the very deep ($A_{\rm cell}\sim 100$) region of the outer crust. Assuming a 100\% limit for the maximum relative increase in pressure due to spin-down, only a few of the deepest shells can reach the onset of neutron drip; see Table~\ref{tab:pacoc3}. 

The estimation of the rotation frequency of a newly born NS that is not spun up by accretion is a complicated task depending on many assumptions. The analysis of observed pulsars gives initial rotation periods between $10$ and several hundreds of milliseconds \citep{Faucher2006}; these results can be supported by theoretical supernova modeling \citep[see][]{Janka2022}. Even for the lower limit of this range ($P\sim 10-20$ ms), the relative increase in pressure after slowing down would be less than 1\%. A small increase like this would not trigger any exothermic reaction in the catalyzed crust. This is not the case for a fully accreted crust when reactions occur continuously with the pressure increase \citep{Gusakov2015}. 

\subsection{Temporary and permanent heat sources}

To compare the modeling of a partially accreted crust with the observed luminosity of some soft quiescent X-ray transients, a detailed catalog of the heat sources must be assembled. The dynamic nature of the crust compression suggests that this catalog should be presented as a function of time, especially because the thermal relaxation of NSs is our observable of interest. However, to keep the catalog independent of the binary accretion rate or the spin slowdown (given the rotation period ${\cal P}$ and its time derivative $\dot{\cal P}$), the catalog was established as a function of the compression $\Delta P$. To implement this catalog in the relativistic equations of cooling and transport, $\dot{M}(t)$ or ${\dot{\cal P}}(t)$ must be specified. For partially accreted crusts, we propose the following equation:
\begin{align}\label{eq:compressionAccretion}
    \Delta P &= 2.1 \times 10^{24} \frac{\dot{M}_{-10} t_{\rm acc}/{\rm yr}}{\alpha_*}, 
\end{align}
where $t_{\rm acc}$ is the time elapsed since accretion started. We refer to \citet{IidaSato1997} for the relation between the relative increase in pressure and the rotation frequency decrease of the star. 

\begin{figure}
\resizebox{\hsize}{!}{\includegraphics{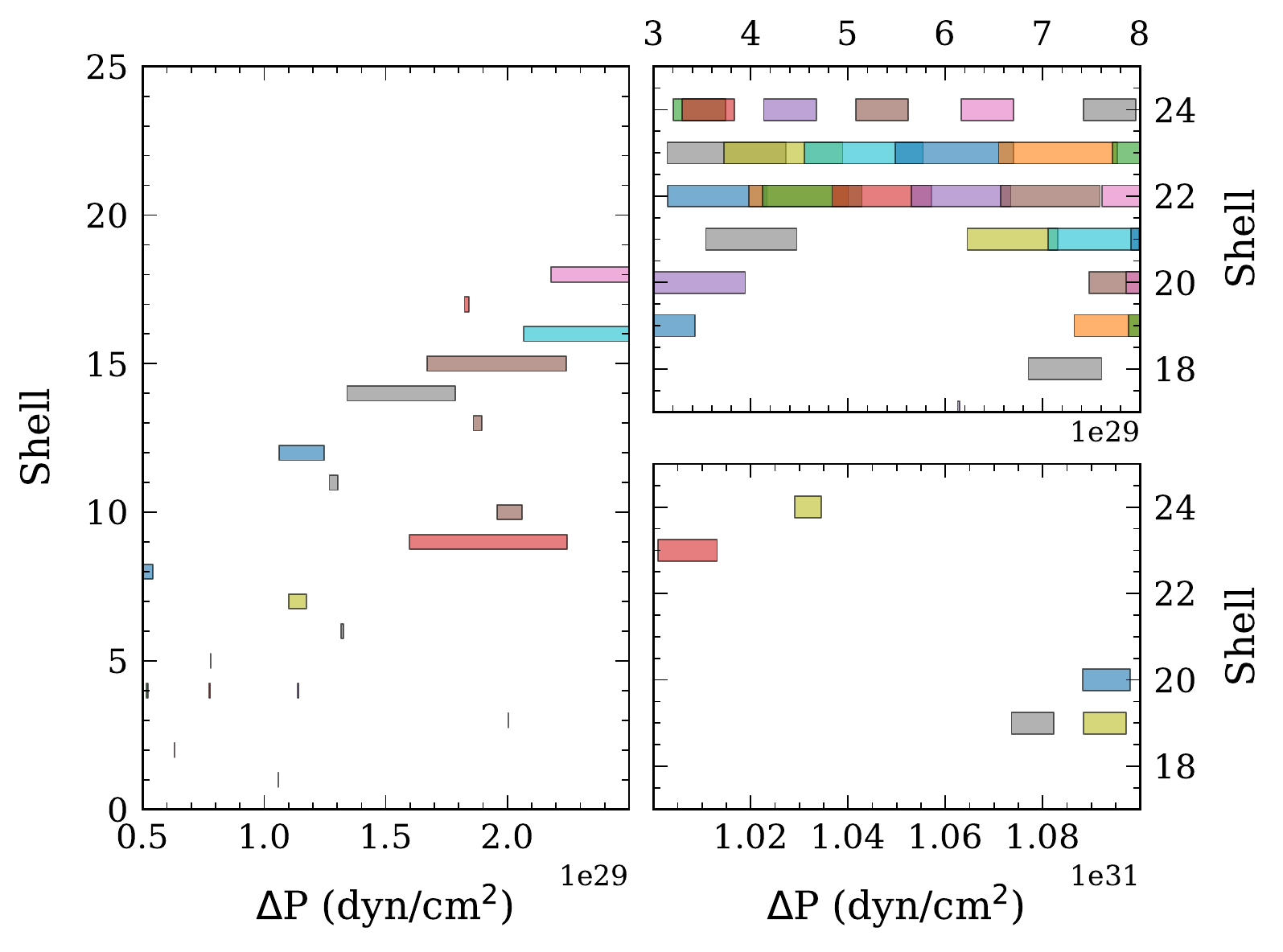}}
\caption{Heating activity expressed in terms of the compression $\Delta P$ required for heat sources to transform each shell of the originally catalyzed compressed outer crust from  the parent to the daughter nuclei. The colors have no special meaning other than to distinguish between overlapping reactions in a shell. For details, see the text.}
\label{fig:lasting}
\end{figure}

As opposed to the fully accreted crust approximation and the sources of the accreted material as part of the partially accreted crust, the sources of the compressed crust are of temporary character in both time and position: they appear, then disappear at different depths of the crust. This is presented in Fig.~\ref{fig:lasting} for the 24 shells of the originally catalyzed compressed outer crust at various values of the compression. Temporary as they are, the sources of the originally catalyzed compressed outer crust last for a certain compression range: the exothermic reaction at the origin of the heat source first appears in the deepest part of the shell and then invades the shell progressively until the entire shell $i$ has been transformed from nuclei $(A_{\rm cell}, Z, N)$ (parent nuclei) to $(A^{'}_{\rm cell}, Z^{'}, N^{'})$ (daughter nuclei). In the left plot of Fig.~\ref{fig:lasting}, for ${\Delta P=[0.5-1]10^{29}\;{\rm dyn/cm^2}}$, heat sources occur in the shallowest shells, which are thin: there are periods of heat extinction. For deep shells, which present a wide pressure range $\big( P^i_{\rm bot} - P^i_{\rm top} \big)$, one exothermic reaction sometimes has not yet filled the entire shell before another starts at the bottom end $P^i_{\rm bot}$, which results in an overlap of reactions within one shell. The shallowest shells have narrow pressure ranges. It takes less compression to fill them with daughter nuclei, and they allow for more lull than in the deepest shells. In the upper right plot, there is no extinction because a reaction proceeds in at least one shell. For the highest values of the compression, there are fewer heat sources, as presented in the lower right plot of Fig.~\ref{fig:lasting}, and quiescence exists.

The results for the heat sources of the partially accreted crust are presented in a movie (\textit{movHeat.png}) available in Ancillary files  : heat sources per nucleon for both the accreted material part of the partially accreted crust and the originally catalyzed compressed outer crust are given as a function of the pressure in the crust at each $\Delta P$ on the upper figure of the movie. The temporary nature of sources in the originally catalyzed compressed outer crust stands out, whereas the sources in the accreted material part settle permanently.

\section{Properties of the compressed outer crust}\label{sec:properties}

\subsection{Timescales}
Each (astrophysical or nuclear) process discussed in this paper is characterized by a typical timescale. We briefly present the relevant timescales for the different phenomena.

For a star in hydrostatic equilibrium, the pressure and gravitation counterbalance each other. If an imbalance between the two occurs in a given region of the crust, the NS structure (to a good approximation, this concerns only the structure of the crust)  adjusts on  a timescale called dynamical  in order to cancel this imbalance, and reaches a new hydrostatic equilibrium. This timescale is denoted $\tau_{\rm dyn}$ and is about $0.1~{\rm ms}$; it is the time needed for a sound wave to cross the crust.

In the context of accretion, a piece of matter located at a pressure $P_{\rm ori}$ in the original crust is pushed to a pressure threshold $P_{\rm th}$ at which a specific reaction takes place within a timescale denoted $\tau_{\rm acc}$,
\begin{equation} \label{eq:timescaleAccretion}
    \tau_{\rm acc} = 4.8 \times 10^5 \frac{\alpha_*}{\langle\dot{M}\rangle_{-10}} \big(P_{\rm th, 30} - P_{\rm ori, 30} \big) \rm \; yr, 
\end{equation}
where $P_{30}$ is the pressure in 10$^{30}$\;dyn/cm$^2$. As presented in Fig.~\ref{fig:lasting}, when a reaction pressure threshold is reached at the bottom of the original shell $i$, the additional compression required to ensure that the entire shell goes through this reaction is directly related to the thickness in pressure of the shell. We note it $\tau_{\rm st}$,
    \begin{equation}\label{eq:shellFilling}
        \tau_{\rm st} = 4.8 \times 10^5  \frac{\alpha_*}{\langle\dot{M}\rangle_{-10}}(P^{i}_{\rm bot,30} - P^{i}_{\rm top,30}) {\rm yr}.
    \end{equation}

The lifespan of the partially accreted crust, in other words, the timescale of pushing the crust to the core, denoted $\tau_{\rm pac}$, is given by 
\begin{equation}
    \tau_{\rm pac} = 5 \times 10^7\frac{\alpha_*}{\langle\dot{M}\rangle}_{-10} \;{\rm yr}.
\end{equation}
    
Assuming  dipolar magnetic model for a pulsar of rotation period $\mathcal{P}$, we can give a rough estimation of the pulsar lifetime. For a constant magnetic field $B$ and a moment of inertia $I$, the pulsar becomes nonobservable (dead) after a spin-down time 
\begin{equation}
    \tau_{\rm sd}= 6.5 \times 10^7 \mathcal{P}^2\frac{I_{45}}{R_{6}^6 B_{12}^2} \;{\rm yr},
\end{equation}
with $B_{12}$ the polar magnetic field in 10$^{12}$\;G, and $I_{45}$  the moment of inertia in 10$^{45}$\;g cm$^2$ \citep[see][Sect. 3.2]{IidaSato1997}. 

The rate of electron captures on nuclei per ${\rm cm}^3$ per second and per one nucleon ${\cal R}_{\rm cap}$ is related to the electron capture timescale by $\tau_{\rm cap}=1/{\cal R}_{\rm cap}$. In a steady state of a partially accreted crust, this results from the balance of the accretion flow and the capture rate. In this way, we obtain 
\begin{equation} 
    \tau_{\rm cap}=\frac{{4\pi R^4 P_{\rm th}}}{{G\dot{M}M}} \approx 10^5~\frac{{P_{\rm th,30} (R_6)^4}}{{\langle\dot{M}\rangle_{-10}(M/{\rm M}_\odot)}}~{\rm yr}~.
\end{equation}
On the other hand, if we consider nonequilibrium electron captures in a partially accreted crust after accretion has stopped, the capture timescale for the reaction $(Z=26,N=30)$ $\to$ $(Z=28,N=28)$ is about 10\;yr. 

The threshold for pycnonuclear reactions in a given shell should be determined by comparing two timescales, the pycnonuclear $\tau_{\rm pyc}$ , and the timescale of accretion $\tau_{\rm acc}$. Pycnonuclear fusion switches on as soon as $\tau_{\rm pyc} < \tau_{\rm acc}$ \citep[see][]{Haensel2003}. However, the uncertainties in the pycnonuclear reaction rates are very large \citep{Yakovlev2006}. Fortunately, because the function $\tau_{\rm pyc}(P)$ is very steep, this uncertainty does not significantly change the location of the energy sources and the total energy release due to the pycnonuclear reactions \citep{Haensel2008}.

\begin{figure}
\resizebox{\hsize}{!}{
\begin{tikzpicture}
\draw[->, line width=1pt] (0,-1) -- (0,5);
\draw (0.35,4.7) node[above]{$r$};

\fill[gray!40] (0.5,-1) rectangle (8,0);
\draw (4.25,-0.7) node[above]{Shell $i+2$};

\fill[gray!40] (0.5,0) rectangle (8,0.4);
\draw[->,line width=2pt][gray!60] (2,0.4) -- (2,0.7);
\draw[->,line width=2pt][gray!60] (6.5,0.4) -- (6.5,0.7);
\draw[line width=1pt] (0.5,0) -- (8,0);
\draw (4.25,0.8) node[above]{Shell $i+1$};
\draw[->,line width=1pt][red!60] (2,2) -- (2,1.4);
\draw[->,line width=1pt][red!60] (3.5,2) -- (3.5,1.4);
\draw[->,line width=1pt][red!60] (5,2) -- (5,1.4);
\draw[->,line width=1pt][red!60] (6.5,2) -- (6.5,1.4);

\fill[gray!40] (0.5,2) rectangle (8,3.5);
\draw[->,line width=2pt][gray!60] (2,3.5) -- (2,3.8);
\draw[->,line width=2pt][gray!60] (6.5,3.5) -- (6.5,3.8);
\draw[line width=1pt] (0.5,2) -- (8,2);
\draw (4.25,2.7) node[above]{Shell $i$};

\draw[line width=1pt]  (0.5,4) -- (8,4);
\draw (4.25,4.2) node[above]{Shell $i-1$};

\end{tikzpicture}
}
\caption{Schematic of the neutron-drip anomaly. Areas of the shells with free neutrons are represented in gray.}
\label{fig:schematicND}
\end{figure}
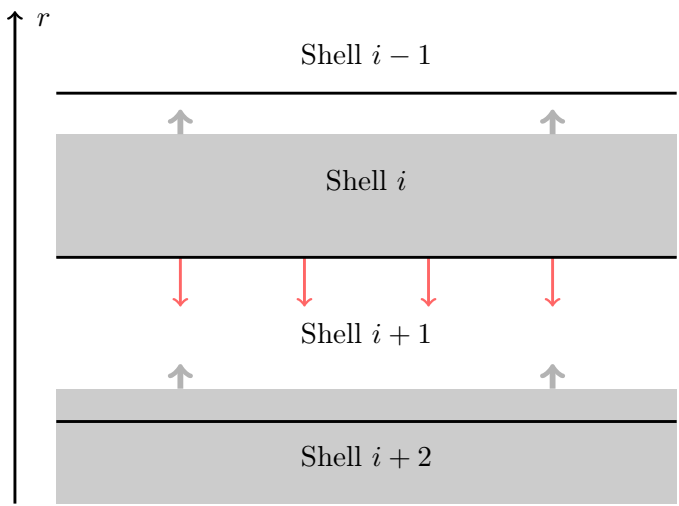

\subsection{Neutron-drip anomaly}
\label{sect:n-drip-anomaly}

For a compressed crust, the neutron-drip point depends on the original shell $(A_{\rm cell},Z,N)$. The configuration displayed in Fig.~\ref{fig:schematicND} can occur at compression between $8 \times 10^{29}$ and $1.5 \times 10^{30}$\;dyn/cm$^2$: the shallower shell $i$ presents a layer with free neutrons, and its deeper neighbor, shell $i+1$ does not. In this case, shell $i$ has started neutron drip at an earlier stage of compression than shell $i+1$. We are left with a layer in shell $i+1$ without free neutrons surrounded by two layers with free neutrons. This configuration emerges due to the compression of different shells, therefore it cannot appear in a fully accreted crust approximation, nor in the accreted material part of the partially accreted crust.

For example, in the single nucleus model, after an increase in pressure ${\Delta P \sim 10^{30}\;{\rm dyn/cm^2}}$, (equivalently ${\sim 2.5 \times 10^{-5}\;{\rm M}_{\odot}}$ accreted on the surface of the original crust, see Eq.\eqref{eq:compressionAccretion} for a 1.4\;M$_{\odot}$ star), neutron drip occurs at the bottom of the 14th$^{\rm }$ shell when it has not yet in the deepest end of shell 15. This neutron-drip anomaly can be tracked in Fig.~\ref{fig:NDcompression}: from the deepest to the shallowest shell, the global trend of the neutron-drip onset is from left (lowest compression) to right (highest compression). When two neighbor shells do not follow this trend, the anomaly occurs. In the continuous approach, the anomaly does occur, but not necessarily for the same shells as in the single-nucleus model. 

As the scheme of Fig.~\ref{fig:schematicND} suggests, the timescale described in Eq.~\eqref{eq:shellFilling} entails that if shell $i$ (shallower) starts neutron drip (at its deepest end) before shell $i+1$ (deeper) has ended emitting  free neutrons, the layers of these shells will present an anomalous configuration. This is the case for shells 21 and 22 in the continuous approach: in Fig.~\ref{fig:NDcompression}, shell 22 starts dripping at relative compression $\delta P = 17$\%, while shell 21 starts at $\delta P = 29$\%, which is in accordance with the neutron-drip trend. However, the additional compression required for the reaction to go through the entire shell after starting at its bottom implies that shell 22 will finish filling up with free neutrons at $\delta P= 35$\%, that is, well after shell 21 starts dripping.

For certain values of the compression, this phenomenon is repeated several times: at $\Delta P \simeq 9\times 10^{29}$\;dyn/cm$^2$ (in the single nucleus model), no shell of the originally catalyzed compressed outer crust has started to drip; at $\Delta P \simeq 1.56\times 10^{30}$\;dyn/cm$^2$, all shells have been pushed to the inner crust. For $\Delta P \simeq 1.23\times 10^{30}$\;dyn/cm$^2$, six layers are involved in the anomaly, thus alternating layers with and without free neutrons. After $\Delta P \simeq 1.6 \times 10^{30}$\;dyn/cm$^2$ in the single-nucleus model, all shells of the originally catalyzed compressed outer crust have undergone neutron drip and have been pushed to depth corresponding to the inner crust; in the continuous approach, after compression $\Delta P \simeq 1.1 \times 10^{30}$\;dyn/cm$^2$, the whole outer crust has been pushed to the inner crust.

Up to now, we considered a quasi-static scenario for an anomalous neutron drip. From now on, the kinetics of dripped neutrons are included. We considered the emergence of a new layer. In the following, we focus on a layer of daughter nuclei from two successive electron captures, but the same reasoning can be applied to pycnonuclear fusion. The emitted neutrons scatter on nuclei via strong interaction; scattering on electrons can  be neglected because it is due to  magnetic moments coupling only. Under the action of the gravitational field acceleration $g$, neutrons drift downward. Their flow velocity is denoted by $V_n$ and the neutron number current by $j_n=n_n V_n$, where $n_n$ is neutron number density. The  balance of the gravitational force $m_n g$  with a resistance force from the scattering of neutrons on nuclei results in the relation
\begin{equation}
    V_n=b m_n  g~, 
\end{equation}
with $b$ the mobility parameter of the neutron gas in the lattice of nuclei, as presented in \citet{pitaevskii2012}. Nuclei are approximated by hard spheres of radius ${r_A=1.2\; A^{1/3}\;{\rm fm}}$, with $1.2$\;fm being the radius of one nucleon; the neutron-nucleon scattering is treated as elastic. Under these assumptions, the transport cross-section for neutron scattering is $\sigma_{\rm t}=\pi r_A^2$. The scatterer's mass, that is to say, the mass of nuclei with $A\simeq 100$, is two orders of magnitude higher than the neutron mass $m_n$. Moreover, the number density of the neutron gas $n_n$ is smaller than the number density of nuclei $n_A$ (see below). At a prevailing temperature of $\sim 10^8~{\rm K}$, neutron gas is nondegenerate, therefore we can apply the method described in \citet{pitaevskii2012}, in which the diffusion of the Boltzmann gas of light particles in the gas of heavy particles approximated by hard elastic spheres was considered. Einstein's relation is also used between the diffusion coefficient $D$ and mobility $b$, $D=k_{\rm B}Tb$ as in \citet{Landau1987Fluid}; $k_B$ is the Boltzmann constant. 

In our case, neutrons drift downward under the action of an external force $f=m_n g$, and the formula for drift velocity \citep{pitaevskii2012} is 
\begin{equation}
    V_n=    {{g m_n v^n_T}\over{ 3 n_A\sigma_{\rm t} k_{\rm B} T}}~,
\end{equation}
where $v_T^n=\sqrt{k_{\rm B}T/m_n}$ is the mean microscopic (thermal)  speed of neutrons in the neutron gas, 
\begin{equation}
    v^n_T=\sqrt{\frac{k_B T}{m_n}}=9.1\times 10^7 \sqrt{\frac{T}{10^8~{\rm K}}}~{\rm cm/s}~.    
\end{equation}
The  product $n_A \sigma_{\rm t}$ is  the inverse transport mean free path of  neutrons, $\lambda_{\rm t}^{-1}$, such that 
\begin{equation}
    \lambda_{\rm t} = 1.705\times 10^{-9}~ \frac{(A_{100})^{1/3}}{\rho_{11}}~{\rm cm}, 
\end{equation}
with $\rho_{11}$ the density in ${10^{11}\;{\rm g/cm^3}}$, and A$_{100}$ the nucleon number divided by 100. For a typical ${k_{\rm B}T=6.905\times 10^{-8} T/(5\times 10^8\;{\rm K})\;{\rm erg}}$, we obtain for the drift velocity 
\begin{equation}\label{eq:drift}
    V_n= 478~\frac{(A_{100})^{1/3}}{\rho_{11}} \bigg(\frac{T}{10^8~{\rm K}}\bigg)^{-1/2} \frac{g}{g_{*}} \;{\rm m}/ {\rm yr},
\end{equation}
where $g_{*}= 1.3 \times 10^{14} M/M_{\odot} / R_6^2$ cm/s$^2$ is the Newtonian surface gravitational acceleration of the star.

The neutron drift  process was considered in the reference frame associated with lattice of nuclei, which actually moves inwards  due to compression implied by accretion. The radial velocity $V_{\rm c}$ is given by
\begin{equation}\label{eq:radial}
    V_{\rm c}=\frac{\dot{M}}{4\pi R^2 \rho}= 1.5 \frac{\dot{M}_{-10}}{R_6^2 \rho_{11}}\;{\rm mm/yr}~,
\end{equation}
with $\rho_{11}$ the density in $10^{11}$\;g/cm$^3$. $V_{\rm c}$ is negligibly small compared to  $V_n$, therefore, neutrons leave the layer right after their emission and thermal equilibration, and drift downward through the outer crust to the inner crust.

Using the estimates derived above, we can evaluate the number density of the gas of the emitted neutrons: ${n_n=k\cdot n V_{\rm c}/V_n}$ so that the free neutron fraction ${X_n=n_n/n\simeq k\cdot 10^{-7}}$. This  corresponds to a diluted ideal neutron gas, described by Boltzmann statistics, because the condition for the kinetic energy $\varepsilon_p$ of a neutron of momentum $p$, and the chemical potential (without rest energy) $\mu_n$ satisfies ${\rm e}^{(\mu_n-\varepsilon_p)/k_{\rm B}T}\ll 1$ \citep{landau1980stat}. For an ideal gas of neutrons, the chemical potential is  

\begin{equation}
    \mu_n =k_{\rm B} T {\rm ln} \left[n_n \left( {2\pi\hbar^2} \over m_n k_{\rm B} T\right)^{3/2}\right]~,
\end{equation}
therefore a  sufficient condition for the validity of Boltzmann statistics is that the dimensionless parameter ${n_n\left({2\pi\hbar^2}/m_n k_{\rm B}T\right)^{3/2}\ll 1}$. We find that at density $\sim 10^{11}$\;g/cm$^3$, this dimensionless parameter for neutron gas is indeed very small ($\sim 10^{-3}$).

Considering neutron number conservation and hydrostatic equilibrium, we can estimate the thickness of the layer of neutron emission, or equivalently, of electron capture. We denote with $W$ the reaction threshold for the electron capture and with $\mu_e$ the chemical potential of electrons. We introduce the dimensionless variable 
\begin{equation}
    y = \frac{\mu_e - W}{W}.
\end{equation}
An element of matter is pushed to the pressure of electron capture threshold $P_0$, at which $\mu_e = W$. If this piece of matter is compressed further (pushed deeper in the star), $y$ increases, but can be considered small ($y \ll 1$). The pressure is mainly contributed by ultrarelativistic electrons. Above the threshold $P_0$ , it can be approximated by
\begin{equation}
    P \simeq P_0 (1 + 4y).
\end{equation}
Assuming allowed beta transition, the reaction rate of the electron capture is proportional to $y^3$. From the relativistic equations for hydrostatic equilibrium, 
\begin{equation}
    y = \frac{g \rho z}{4P_0},
\end{equation}
with $z$ the depth relative to the capture threshold location and $g$ the surface gravity acceleration ($\sim 10^{14}$\;cm/s$^2$). Electron captures with neutron emission involved in the neutron-drip anomaly occur around $\rho \simeq 10^{11}$\;g/cm$^3$, and $P\simeq 10^{30}$\;dyn/cm$^2$. From the drift and radial velocity estimated in Eqs.~\eqref{eq:drift} and \eqref{eq:radial} and from the threshold energy $W\simeq 20$ MeV, we obtain at the bottom of the electron capture (neutron emission) layer $y\simeq 10^{-5}$ and a thickness of the layer $\simeq 3$~cm.

On the one hand, the shells from which free neutrons escape have a lower nucleon number per cell: with the example of shells 14 and 15 at $\Delta P\simeq 10^{30}$\;dyn/cm$^2$, after neutron drip, the daughter nucleus (102,98,30) with four free neutrons is present at pressure $P= 1.2\times 10^{30}$\;dyn/cm$^2$. This nucleus, after losing its four free neutrons that drifted downward, is stable at this pressure. On the other hand, the shells into which the free neutrons have drifted will present a higher neutron number density. Overall, the composition of the compressed crust must be reevaluated. However, treating the interaction of drifting neutrons with deeper shells in the crust requires considerations outside the single-nucleus approach, which is beyond the scope of the present paper.

\subsection{Density-related instability}\label{sec:drinstablity}

Each exothermic reaction triggered at the bottom of one shell of the originally catalyzed compressed outer crust is accompanied with a  density jump. After this reaction, the density of the layer just below this shell may be lower. This density inversion can be subject to the Rayleigh-Taylor instability applied to a lattice, also referred to as the elastic Rayleigh-Taylor instability. The study of this type of instability indicates that it develops when the density jump is larger than 10\% \citep{Blaes1990}. The fulfillment of this condition is represented with yellow stars in the movie (\textit{movHeat.png}) available in Ancillary files. From the movie, it can be concluded that these potentially unstable configurations appear for an increase in pressure in the range ${[8 \times 10^{29} - 3 \times 10^{30}]\;{\rm dyn/cm^2}}$, and occur after accretion of ${\sim 10^{-4}\;{\rm M}_{\odot}}$.

To estimate the energy available due to the swapping of the layers subject to elastic Rayleigh-Taylor instability, we used the formula in \citet{Blaes1990},
\begin{align}
     \Delta E =& \frac{4\pi R^4}{GM} \big(P_{1, \rm bot} - P_{1, \rm top} \big) \big(P_{2, \rm bot} - P_{2, \rm top} \big) \bigg( \frac{1}{\rho_2} - \frac{1}{\rho_1}\bigg)\nonumber \\
     \simeq & 10^{44} \frac{R_{6}^4}{M/M_{\odot}} \big(P_{1, \rm bot} - P_{1, \rm top} \big)_{28} \nonumber\\
     & \times \big(P_{2, \rm bot} - P_{2, \rm top} \big)_{28} \bigg( \frac{1}{\rho_{2,11}} - \frac{1}{\rho_{1,11}} \bigg) {\rm erg} , 
\end{align}
with $\big(P_{i, \rm bot} - P_{i, \rm top} \big)_{28}$ and $\rho_{i,11}$  the pressure thickness in ${10^{28}\;{\rm dyn/cm^2}}$ and mean density in ${10^{11}\;{\rm g/cm^3}}$, respectively, of the layer $i$ ($i=1,2$).
The pressure range of the layers subject to elastic Rayleigh-Taylor instability is typically $10^{28}$ or $10^{29}$\;dyn/cm$^2$. This corresponds to a layer mass of ${\sim 10^{-6}\;{\rm M}_{\odot}}$ and to a timescale of the formation of these layers of ${\sim 10^4\;{\rm yr}/\dot{M}_{-10}}$.

\section{Conclusion}
We studied the evolution of the structure, the composition, and the heat sources of a neutron star crust during the first stage of accretion, when the amount of matter accreted is smaller than the mass of the crust and the originally catalyzed crust has not been replaced by the accreted material. The equation of state of the catalyzed outer crust, of the accreted material, and of the original crust that is compressed under the pressure of accreted material has been established using a compressible liquid-drop model. 
The catalyzed outer crust consists of 24 shells (in the single-nucleus model), containing nuclei with mass numbers $A$ between 56 and 132, eleven of  which are constrained by laboratory measurements.
We reconstructed part of the hybrid crust (accreted material part, and the compressed originally catalyzed outer crust) as a function of the compression by accreted material.

Two types of compression-related exothermic reactions occur in the hybrid crust: electron captures (in the inner crust, accompanied by neutron emissions), and pycnonuclear fusions. For a compression limit of around 2$\times 10^{32}$\;dyn/cm$^2$, the originally catalyzed compressed outer crust hosts around 170 different nuclei. Exothermic reactions leading to this variety of nuclei deposit heat in the compressed outer crust. The largest heat per nucleon is released in the shallowest shells of the original crust. However, most of the total heat is released in the deepest shells. We conclude that during the initial stage of accretion, the heat sources of the original crust that is compressed are not negligible compared to that of the accreted material part of the crust. Moreover, the composition of the originally catalyzed compressed outer crust is significantly different from that of the fully accreted crust. This must be taken into consideration for the transport properties of the crust, and it should not be overlooked when slowly accreting neutron star cooling is modeled.

Compression of the crust may occur for a spinning-down neutron star. Exothermic reactions can be triggered for the deepest layers of the outer crust, when the relative compression of $\sim 50$\% of the crust shell is evaluated. 

The compression of the original outer crust into the inner crust leads to a neutron-drip anomaly. The neutron-drip point of the 24 shells of the originally catalyzed outer crust was calculated with the single-nucleus approach as well as with the continuous approach presented in \citet{Chamel2015}. In both cases, the compression of the originally catalyzed outer crust into the inner crust led to a configuration in which free neutrons may be found above a shell that has not reached its neutron-drip point. After evaluating the kinetics of free neutrons, we conclude that they are emitted in thin layers and will drift deeper into the star nearly instantaneously. The loss of neutrons per cell leads us to conclude that the composition of the shells concerned with this anomaly must be reevaluated, but this is beyond the scope of the current work. Potentially, the compressed catalyzed crust can be subject to the Rayleigh-Taylor instability; the existence of regions with density inversion is a quite common feature of our model. However, this problem needs additional studies including the timescale of the formation of the inverse density layers, and development of the instability.

\section{Acknowledgments}

We thank N. Chamel and A. F. Fantina for useful discussions. We are grateful to the referee for constructive input. The authors acknowledge the financial support of the National Science Center, Poland grant 2018/29/B/ST9/02013 (LS, JLZ and PH) and 2017/26/D/ST9/00591 (MF and LS).


\begin{thebibliography}{41}
\expandafter\ifx\csname natexlab\endcsname\relax\def\natexlab#1{#1}\fi

\bibitem[{{Baglio} {et~al.}(2016){Baglio}, {D’Avanzo}, {Campana}, {Goldoni},
  {Masetti}, {Muñoz-Darias}, {Patiño-Álvarez}, \& {Chavushyan}}]{Baglio2016}
{Baglio}, M.~C., {D’Avanzo}, P., {Campana}, S., {et~al.} 2016, \aap, 587,
  A102

\bibitem[{{Bildsten}(1998)}]{Bildsten1998}
{Bildsten}, L. 1998, in NATO Advanced Study Institute (ASI) Series C, Vol. 515,
  The Many Faces of Neutron Stars., ed. R.~{Buccheri}, J.~{van Paradijs}, \&
  A.~{Alpar}, 419

\bibitem[{{Blaes} {et~al.}(1990){Blaes}, {Blandford}, {Madau}, \&
  {Koonin}}]{Blaes1990}
{Blaes}, O., {Blandford}, R., {Madau}, P., \& {Koonin}, S. 1990, \apj, 363, 612

\bibitem[{{Blaschke} \& {Chamel}(2018)}]{Blaschke2018}
{Blaschke}, D. \& {Chamel}, N. 2018, Phases of Dense Matter in Compact Stars,
  ed. L.~{Rezzolla}, P.~{Pizzochero}, D.~I. {Jones}, N.~{Rea}, \&
  I.~{Vida{\~{n}}a} (Cham: Springer International Publishing), 337--400

\bibitem[{{Bonanno} \& {Urpin}(2015)}]{Bonanno2015}
{Bonanno}, A. \& {Urpin}, V. 2015, \aap, 574, A63

\bibitem[{{Brown} {et~al.}(1998){Brown}, {Bildsten}, \& {Rutledge}}]{Brown1998}
{Brown}, E.~F., {Bildsten}, L., \& {Rutledge}, R.~E. 1998, \apj, 504, L95–L98

\bibitem[{{Brown} \& {Cumming}(2009)}]{Brown_2009}
{Brown}, E.~F. \& {Cumming}, A. 2009, \apj, 698, 1020–1032

\bibitem[{{Chamel} {et~al.}(2021){Chamel}, {Fantina}, {Suleiman}, {Zdunik}, \&
  {Haensel}}]{chamel2021}
{Chamel}, N., {Fantina}, A.~F., {Suleiman}, L., {Zdunik}, J.-L., \& {Haensel},
  P. 2021, {Universe}, 7, 193

\bibitem[{{Chamel} {et~al.}(2015){Chamel}, {Fantina}, {Zdunik}, \&
  {Haensel}}]{Chamel2015}
{Chamel}, N., {Fantina}, A.~F., {Zdunik}, J.~L., \& {Haensel}, P. 2015, \prc,
  91, 055803

\bibitem[{Chamel \& Haensel(2008)}]{Chamel2008}
Chamel, N. \& Haensel, P. 2008, Living Reviews in Relativity, 11

\bibitem[{{Degenaar}(2015)}]{Degenaar2015}
{Degenaar}, N. 2015, {Probing the physics of neutron stars using Terzan 5},
  Chandra Proposal

\bibitem[{{Degenaar} {et~al.}(2015){Degenaar}, {Wijnands}, {Bahramian}, R.,
  {Heinke}, {Brown}, {Fridriksson}, {Homan}, {Cackett}, {Cumming}, {Miller},
  {Altamirano}, \& {Pooley}}]{Degenaar2015-2}
{Degenaar}, N., {Wijnands}, R., {Bahramian}, A., {et~al.} 2015, Neutron star
  crust cooling in the Terzan 5 X-ray transient Swift J174805.3-244637

\bibitem[{{Faucher-Gigu{\`e}re} \& {Kaspi}(2006)}]{Faucher2006}
{Faucher-Gigu{\`e}re}, C.-A. \& {Kaspi}, V.~M. 2006, \apj, 643, 332

\bibitem[{Fortin {et~al.}(2010)Fortin, Grill, Margueron, Page, \&
  Sandulescu}]{Fortin2010}
Fortin, M., Grill, F., Margueron, J., Page, D., \& Sandulescu, N. 2010, Phys.
  Rev. C, 82, 065804

\bibitem[{Fortin {et~al.}(2018)Fortin, Taranto, Burgio, Haensel, Schulze, \&
  Zdunik}]{Fortin2018}
Fortin, M., Taranto, G., Burgio, G.~F., {et~al.} 2018, \mnras, 475, 5010–5022

\bibitem[{{Gupta} {et~al.}(2007){Gupta}, {Brown}, {Schatz}, {M{\"o}ller}, \&
  {Kratz}}]{Gupta2007}
{Gupta}, S., {Brown}, E.~F., {Schatz}, H., {M{\"o}ller}, P., \& {Kratz}, K.-L.
  2007, \apj, 662, 1188

\bibitem[{{Gupta} {et~al.}(2008){Gupta}, {Kawano}, \& {M{\"o}ller}}]{Gupta2008}
{Gupta}, S.~S., {Kawano}, T., \& {M{\"o}ller}, P. 2008, \prl, 101, 231101

\bibitem[{{Gusakov} {et~al.}(2015){Gusakov}, {Kantor}, \&
  {Reisenegger}}]{Gusakov2015}
{Gusakov}, M.~E., {Kantor}, E.~M., \& {Reisenegger}, A. 2015, \mnras, 453, L36

\bibitem[{{Haensel} \& {Zdunik}(2003)}]{Haensel2003}
{Haensel}, P. \& {Zdunik}, J.~L. 2003, \aap, 404, L33

\bibitem[{{Haensel} \& {Zdunik}(2008)}]{Haensel2008}
{Haensel}, P. \& {Zdunik}, J.~L. 2008, \aap, 480, 459

\bibitem[{{Iida} \& {Sato}(1997)}]{IidaSato1997}
{Iida}, K. \& {Sato}, K. 1997, \apj, 477, 294

\bibitem[{{Janka} {et~al.}(2022){Janka}, {Wongwathanarat}, \&
  {Kramer}}]{Janka2022}
{Janka}, H.-T., {Wongwathanarat}, A., \& {Kramer}, M. 2022, \apj, 926, 9

\bibitem[{Landau \& Lifshitz(1980)}]{landau1980stat}
Landau, L.~D. \& Lifshitz, E.~M. 1980, Statistical {Physics} {Part} {I}, 3rd
  edn. (Amsterdam: Elsevier), bibtex: landa;b;sp80

\bibitem[{{Landau} \& {Lifshitz}(1987)}]{Landau1987Fluid}
{Landau}, L.~D. \& {Lifshitz}, E.~M. 1987, Course of Theoretical Physics,
  Vol.~6, {Fluid Mechanics}, 2nd edn. (Pergamon)

\bibitem[{{Lau} {et~al.}(2018){Lau}, {Beard}, {Gupta}, {Schatz}, {Afanasjev},
  {Brown}, {Deibel}, {Gasques}, {Hitt}, {Hix}, {Keek}, {M{\"o}ller},
  {Shternin}, {Steiner}, {Wiescher}, \& {Xu}}]{Lau2018}
{Lau}, R., {Beard}, M., {Gupta}, S.~S., {et~al.} 2018, \apj, 859, 62

\bibitem[{{Mackie} \& {Baym}(1977)}]{Mackie1977}
{Mackie}, F.~D. \& {Baym}, G. 1977, \nphysa, 285, 332

\bibitem[{Marino {et~al.}(2019)Marino, Del Santo, Cocchi, D’Aì, Segreto,
  Ferrigno, Di Salvo, Malzac, Iaria, \& Burderi}]{Marino2019}
Marino, A., Del Santo, M., Cocchi, M., {et~al.} 2019, \mnras, 490, 2300

\bibitem[{Meisel {et~al.}(2018)Meisel, Deibel, Keek, Shternin, \&
  Elfritz}]{Meisel_2018}
Meisel, Z., Deibel, A., Keek, L., Shternin, P., \& Elfritz, J. 2018, Journal of
  Physics G: Nuclear and Particle Physics, 45

\bibitem[{Oertel {et~al.}(2017)Oertel, Hempel, Kl\"ahn, \& Typel}]{Oertel2017}
Oertel, M., Hempel, M., Kl\"ahn, T., \& Typel, S. 2017, Rev. Mod. Phys., 89,
  015007

\bibitem[{{Parikh} {et~al.}(2013){Parikh}, {Jos{\'e}}, {Sala}, \&
  {Iliadis}}]{Parikh2013}
{Parikh}, A., {Jos{\'e}}, J., {Sala}, G., \& {Iliadis}, C. 2013, Progress in
  Particle and Nuclear Physics, 69, 225

\bibitem[{Parikh {et~al.}(2017)Parikh, Wijnands, Degenaar, Ootes, Page,
  Altamirano, Cackett, Deller, Gusinskaia, Hessels, Homan, Linares, Miller, \&
  Miller-Jones}]{Parikh2017}
Parikh, A.~S., Wijnands, R., Degenaar, N., {et~al.} 2017, \mnras, 466, 4074

\bibitem[{Pearson {et~al.}(2018)Pearson, Chamel, Potekhin, Fantina, Ducoin,
  Dutta, \& Goriely}]{Pearson2018}
Pearson, J.~M., Chamel, N., Potekhin, A.~Y., {et~al.} 2018, \mnras, 481, 2994

\bibitem[{Pitaevskii \& Lifshitz(2012)}]{pitaevskii2012}
Pitaevskii, L. \& Lifshitz, E. 2012, Physical Kinetics: Volume 10 No. vol.~10
  (Elsevier Science)

\bibitem[{{Potekhin} \& {Chabrier}(2021)}]{Potekhin2021}
{Potekhin}, A.~Y. \& {Chabrier}, G. 2021, \aap, 645, A102

\bibitem[{{Sato}(1979)}]{Sato1979}
{Sato}, K. 1979, Progress of Theoretical Physics, 62, 957

\bibitem[{Suvorov \& Melatos(2020)}]{Recycled2020}
Suvorov, A.~G. \& Melatos, A. 2020, \mnras, 499, 3243–3254

\bibitem[{{Wang} {et~al.}(2017){Wang}, {Audi}, {Kondev}, {Huang}, {Naimi}, \&
  {Xu}}]{ame2016}
{Wang}, M., {Audi}, G., {Kondev}, F.~G., {et~al.} 2017, Chinese Physics C, 41,
  030003

\bibitem[{{Wijnands} {et~al.}(2017){Wijnands}, {Degenaar}, \&
  {Page}}]{Wijnands2017}
{Wijnands}, R., {Degenaar}, N., \& {Page}, D. 2017, Journal of Astrophysics and
  Astronomy, 38, 49

\bibitem[{{Yakovlev} {et~al.}(2006){Yakovlev}, {Gasques}, \&
  {Wiescher}}]{Yakovlev2006}
{Yakovlev}, D.~G., {Gasques}, L., \& {Wiescher}, M. 2006, \mnras, 371, 1322

\bibitem[{Zdunik {et~al.}(2016)Zdunik, Fortin, \& Haensel}]{Zdunik2016}
Zdunik, J.~L., Fortin, M., \& Haensel, P. 2016, Neutron star properties and the
  equation of state for its core

\bibitem[{{Zdunik} {et~al.}(2001){Zdunik}, {Haensel}, \&
  {Gourgoulhon}}]{Zdunik2001}
{Zdunik}, J.~L., {Haensel}, P., \& {Gourgoulhon}, E. 2001, \aap, 372, 535

\end{thebibliography}

\appendix
\begin{appendix}

\begin{table*}
\caption{Reactions triggered in the originally catalyzed compressed outer crust (calculated in the MB model) up to the first neutron emission. Nuclei based on experimentally determined masses from AME2016 are presented in bold. The neutron-drip points calculated in the continuous approach are presented in blue. In order of columns, we give the pressure $P$ in (\;dyn/cm$^2$), the density $\rho_{\rm ini}$ right before the reaction, $\lambda =  \Delta \rho / \rho$ the relative change in density due to the reaction in percent, the reaction concerned, the energy per nucleon $Q$ of the reaction, the compression $\Delta P$ at which the reaction is triggered, and the relative compression $\delta P /P$ in percent at which the reaction is triggered. The table is separated into 24 parts, one for each shell: the first line (before the dashed lines) includes the top pressure and density of the original shell (first and second column) as well as the nuclei of the shell. A computer-readable version of this table is provided in Ancillary files (see  \textit{ReactionOCCOCSNM.res} and \textit{ReactionOCCOCContinuous.res}}).
\label{tab:pacoc1}
\centering
\begin{tabular}{ccccccc}
\hline 
\hline 
P & $\rho_{\rm ini}$ & $\lambda$ & Reaction & Q (keV/A) & $\Delta$ P & $\delta P/P $ (in \%) \\
$1.442 \times 10^{22}$ & $7.753 \times 10^{5}$ & - & $^{56}$\textbf{Fe} & \multicolumn{3}{c}{Shell 1} \\
\hdashline
$6.468 \times 10^{26}$ & $1.374 \times 10^{9}$ & 8.2 & $^{56}$\textbf{Fe} + 2e$^-$ $\to$ $^{56}$\textbf{Cr} + 2$\nu_e$ & 37.0 & $6.463 \times 10^{26}$ & $1.2 \times 10^{5}$ \\ 
$1.829 \times 10^{28}$ & $1.811 \times 10^{10}$ & 8.9 & $^{56}$\textbf{Cr} + 2e$^-$ $\to$ $^{56}$\textbf{Ti} + 2$\nu_e$ & 41.2 & $1.829 \times 10^{28}$ & $3.4 \times 10^{6}$ \\ 
$1.059 \times 10^{29}$ & $7.365 \times 10^{10}$ & 9.8 & $^{56}$\textbf{Ti} + 2e$^-$ $\to$ $^{56}$Ca + 2$\nu_e$ & 81.7 & $1.059 \times 10^{29}$ & $2.0 \times 10^{7}$ \\ 
$4.747 \times 10^{29}$ & $2.496 \times 10^{11}$ & 10.9 & $^{56}$Ca + 2e$^-$ $\to$ $^{56}$Ar + 2$\nu_e$ & 46.1 & $4.747 \times 10^{29}$ & $8.8 \times 10^{7}$ \\ 
\textcolor{blue}{$1.064 \times 10^{30}$} & \textcolor{blue}{$5.078 \times 10^{11}$} & \textcolor{blue}{4.3} & \textcolor{blue}{$^{56}$Ar + 2 e$^-$ $\to$ $^{52}$S +  4 n + 2 $\nu_e$} & \textcolor{blue}{41.9} & \textcolor{blue}{$1.064 \times 10^{30}$} & \textcolor{blue}{$2.0 \times 10^{8}$} \\
$1.226 \times 10^{30}$ & $5.649 \times 10^{11}$ & 12.1 & $^{56}$Ar + 2e$^-$ $\to$ $^{52}$S + 4 n + 2$\nu_e$ & 35.1 & $1.226 \times 10^{30}$ & $2.3 \times 10^{8}$ \\ 
\hline
$5.441 \times 10^{23}$ & $8.489 \times 10^{6}$ & - & $^{62}$\textbf{Ni} & \multicolumn{3}{c}{Shell 2} \\
\hdashline
$2.468 \times 10^{27}$ & $3.846 \times 10^{9}$ & 7.5 & $^{62}$\textbf{Ni} + 2e$^-$ $\to$ $^{62}$\textbf{Fe} + 2$\nu_e$ & 44.9 & $2.401 \times 10^{27}$ & $3.6 \times 10^{3}$ \\ 
$2.987 \times 10^{28}$ & $2.679 \times 10^{10}$ & 8.2 & $^{62}$\textbf{Fe} + 2e$^-$ $\to$ $^{62}$\textbf{Cr} + 2$\nu_e$ & 44.2 & $2.981 \times 10^{28}$ & $4.4 \times 10^{4}$ \\ 
$4.283 \times 10^{28}$ & $3.796 \times 10^{10}$ & 4.3 & $^{62}$\textbf{Cr} + 1e$^-$ $\to$ $^{62}$V + 1$\nu_e$ & 0.0 & $4.276 \times 10^{28}$ & $6.4 \times 10^{4}$ \\ 
$6.314 \times 10^{28}$ & $5.297 \times 10^{10}$ & 4.5 & $^{62}$V + 1e$^-$ $\to$ $^{62}$Ti + 1$\nu_e$ & 0.0 & $6.308 \times 10^{28}$ & $9.4 \times 10^{4}$ \\ 
$4.735 \times 10^{29}$ & $2.511 \times 10^{11}$ & 9.8 & $^{62}$Ti + 2e$^-$ $\to$ $^{62}$Ca + 2$\nu_e$ & 41.3 & $4.734 \times 10^{29}$ & $7.1 \times 10^{5}$ \\ 
\textcolor{blue}{$1.014 \times 10^{30}$} & \textcolor{blue}{$4.889 \times 10^{11}$} & \textcolor{blue}{3.8} & \textcolor{blue}{$^{62}$Ca + 2 e$^-$ $\to$ $^{58}$Ar +  4 n + 2 $\nu_e$} & \textcolor{blue}{32.2} & \textcolor{blue}{$1.014 \times 10^{30}$} & \textcolor{blue}{$1.5 \times 10^{6}$} \\
$1.170 \times 10^{30}$ & $5.444 \times 10^{11}$ & 10.9 & $^{62}$Ca + 2e$^-$ $\to$ $^{60}$Ar + 2 n + 2$\nu_e$ & 29.5 & $1.170 \times 10^{30}$ & $1.7 \times 10^{6}$ \\ 
\hline
$6.723 \times 10^{25}$ & $2.654 \times 10^{8}$ & - & $^{58}$\textbf{Fe} & \multicolumn{3}{c}{Shell 3} \\
\hdashline
$4.641 \times 10^{27}$ & $6.207 \times 10^{9}$ & 8.2 & $^{58}$\textbf{Fe} + 2e$^-$ $\to$ $^{58}$\textbf{Cr} + 2$\nu_e$ & 43.1 & $4.571 \times 10^{27}$ & $6.5 \times 10^{3}$ \\ 
$4.552 \times 10^{28}$ & $3.718 \times 10^{10}$ & 8.9 & $^{58}$\textbf{Cr} + 2e$^-$ $\to$ $^{58}$Ti + 2$\nu_e$ & 110.6 & $4.545 \times 10^{28}$ & $6.5 \times 10^{4}$ \\ 
$2.003 \times 10^{29}$ & $1.231 \times 10^{11}$ & 9.8 & $^{58}$Ti + 2e$^-$ $\to$ $^{58}$Ca + 2$\nu_e$ & 45.0 & $2.003 \times 10^{29}$ & $2.8 \times 10^{5}$ \\ 
\textcolor{blue}{$6.598 \times 10^{29}$} & \textcolor{blue}{$3.311 \times 10^{11}$} & \textcolor{blue}{7.1} & \textcolor{blue}{$^{58}$Ca + 2 e$^-$ $\to$ $^{56}$Ar +  2 n + 2 $\nu_e$} & \textcolor{blue}{35.4} & \textcolor{blue}{$6.597 \times 10^{29}$} & \textcolor{blue}{$9.4 \times 10^{5}$} \\
$6.892 \times 10^{29}$ & $3.421 \times 10^{11}$ & 10.9 & $^{58}$Ca + 2e$^-$ $\to$ $^{58}$Ar + 2$\nu_e$ & 44.3 & $6.891 \times 10^{29}$ & $9.8 \times 10^{5}$ \\ 
$1.507 \times 10^{30}$ & $6.830 \times 10^{11}$ & 26.4 & $^{58}$Ar + 4e$^-$ $\to$ $^{46}$Si + 12 n + 4$\nu_e$ & 108.6 & $1.507 \times 10^{30}$ & $2.1 \times 10^{6}$ \\ 
\hline
$7.053 \times 10^{25}$ & $2.821 \times 10^{8}$ & - & $^{64}$\textbf{Ni} & \multicolumn{3}{c}{Shell 4} \\
\hdashline
$8.042 \times 10^{27}$ & $9.615 \times 10^{9}$ & 7.5 & $^{64}$\textbf{Ni} + 2e$^-$ $\to$ $^{64}$\textbf{Fe} + 2$\nu_e$ & 39.0 & $7.471 \times 10^{27}$ & $1.3 \times 10^{3}$ \\ 
$5.228 \times 10^{28}$ & $4.208 \times 10^{10}$ & 8.2 & $^{64}$\textbf{Fe} + 2e$^-$ $\to$ $^{64}$\textbf{Cr} + 2$\nu_e$ & 38.9 & $5.171 \times 10^{28}$ & $9.1 \times 10^{3}$ \\ 
$7.795 \times 10^{28}$ & $6.143 \times 10^{10}$ & 4.3 & $^{64}$\textbf{Cr} + 1e$^-$ $\to$ $^{64}$V + 1$\nu_e$ & 0.0 & $7.738 \times 10^{28}$ & $1.4 \times 10^{4}$ \\ 
$1.143 \times 10^{29}$ & $8.538 \times 10^{10}$ & 4.5 & $^{64}$V + 1e$^-$ $\to$ $^{64}$Ti + 1$\nu_e$ & 0.0 & $1.138 \times 10^{29}$ & $2.0 \times 10^{4}$ \\ 
\textcolor{blue}{$6.487 \times 10^{29}$} & \textcolor{blue}{$3.284 \times 10^{11}$} & \textcolor{blue}{6.4} & \textcolor{blue}{$^{64}$Ti + 2 e$^-$ $\to$ $^{62}$Ca +  2 n + 2 $\nu_e$} & \textcolor{blue}{20.8} & \textcolor{blue}{$6.481 \times 10^{29}$} & \textcolor{blue}{$1.1 \times 10^{5}$} \\
$6.674 \times 10^{29}$ & $3.355 \times 10^{11}$ & 9.8 & $^{64}$Ti + 2e$^-$ $\to$ $^{64}$Ca + 2$\nu_e$ & 39.8 & $6.669 \times 10^{29}$ & $1.2 \times 10^{5}$ \\ 
$1.418 \times 10^{30}$ & $6.493 \times 10^{11}$ & 10.7 & $^{64}$Ca + 2e$^-$ $\to$ $^{60}$Ar + 4 n + 2$\nu_e$ & 42.4 & $1.418 \times 10^{30}$ & $2.5 \times 10^{5}$ \\ 
\hline
$5.725 \times 10^{26}$ & $1.375 \times 10^{9}$ & - & $^{66}$\textbf{Ni} & \multicolumn{3}{c}{Shell 5} \\
\hdashline
$2.261 \times 10^{28}$ & $2.152 \times 10^{10}$ & 7.5 & $^{66}$\textbf{Ni} + 2e$^-$ $\to$ $^{66}$\textbf{Fe} + 2$\nu_e$ & 49.6 & $2.195 \times 10^{28}$ & $3.3 \times 10^{3}$ \\ 
$7.860 \times 10^{28}$ & $5.893 \times 10^{10}$ & 8.2 & $^{66}$\textbf{Fe} + 2e$^-$ $\to$ $^{66}$Cr + 2$\nu_e$ & 79.1 & $7.794 \times 10^{28}$ & $1.2 \times 10^{4}$ \\ 
$3.254 \times 10^{29}$ & $1.852 \times 10^{11}$ & 8.9 & $^{66}$Cr + 2e$^-$ $\to$ $^{66}$Ti + 2$\nu_e$ & 38.6 & $3.247 \times 10^{29}$ & $4.9 \times 10^{4}$ \\ 
\textcolor{blue}{$7.992 \times 10^{29}$} & \textcolor{blue}{$3.962 \times 10^{11}$} & \textcolor{blue}{6.5} & \textcolor{blue}{$^{66}$Ti + 2 e$^-$ $\to$ $^{64}$Ca +  2 n + 2 $\nu_e$} & \textcolor{blue}{16.6} & \textcolor{blue}{$7.986 \times 10^{29}$} & \textcolor{blue}{$1.2 \times 10^{5}$} \\
$9.018 \times 10^{29}$ & $4.338 \times 10^{11}$ & 9.8 & $^{66}$Ti + 2e$^-$ $\to$ $^{66}$Ca + 2$\nu_e$ & 38.4 & $9.011 \times 10^{29}$ & $1.4 \times 10^{5}$ \\ 
$1.556 \times 10^{30}$ & $7.179 \times 10^{11}$ & 23.2 & $^{66}$Ca + 4e$^-$ $\to$ $^{54}$S + 12 n + 4$\nu_e$ & 102.4 & $1.556 \times 10^{30}$ & $2.4 \times 10^{5}$ \\ 
\hline
$6.610 \times 10^{26}$ & $1.561 \times 10^{9}$ & - & $^{86}$\textbf{Kr} & \multicolumn{3}{c}{Shell 6} \\
\hdashline
$9.884 \times 10^{27}$ & $1.180 \times 10^{10}$ & 5.7 & $^{86}$\textbf{Kr} + 2e$^-$ $\to$ $^{86}$\textbf{Se} + 2$\nu_e$ & 29.2 & $8.236 \times 10^{27}$ & 499 \\ 
$4.723 \times 10^{28}$ & $4.031 \times 10^{10}$ & 6.1 & $^{86}$\textbf{Se} + 2e$^-$ $\to$ $^{86}$\textbf{Ge} + 2$\nu_e$ & 23.2 & $4.558 \times 10^{28}$ & $2.8 \times 10^{3}$ \\ 
$1.333 \times 10^{29}$ & $9.319 \times 10^{10}$ & 6.5 & $^{86}$\textbf{Ge} + 2e$^-$ $\to$ $^{86}$Zn + 2$\nu_e$ & 18.8 & $1.317 \times 10^{29}$ & $8.0 \times 10^{3}$ \\ 
$4.385 \times 10^{29}$ & $2.427 \times 10^{11}$ & 7.0 & $^{86}$Zn + 2e$^-$ $\to$ $^{86}$Ni + 2$\nu_e$ & 28.3 & $4.369 \times 10^{29}$ & $2.7 \times 10^{4}$ \\ 
\textcolor{blue}{$8.422 \times 10^{29}$} & \textcolor{blue}{$4.239 \times 10^{11}$} & \textcolor{blue}{5.0} & \textcolor{blue}{$^{86}$Ni + 2 e$^-$ $\to$ $^{84}$Fe +  2 n + 2 $\nu_e$} & \textcolor{blue}{14.7} & \textcolor{blue}{$8.406 \times 10^{29}$} & \textcolor{blue}{$5.1 \times 10^{4}$} \\
$9.668 \times 10^{29}$ & $4.702 \times 10^{11}$ & 7.5 & $^{86}$Ni + 2e$^-$ $\to$ $^{86}$Fe + 2$\nu_e$ & 28.3 & $9.652 \times 10^{29}$ & $5.9 \times 10^{4}$ \\ 
$1.478 \times 10^{30}$ & $6.956 \times 10^{11}$ & 27.1 & $^{86}$Fe + 6e$^-$ $\to$ $^{66}$Ca + 20 n + 6$\nu_e$ & 104.7 & $1.476 \times 10^{30}$ & $9.0 \times 10^{4}$ \\ 
\hline
$1.653 \times 10^{27}$ & $3.195 \times 10^{9}$ & - & $^{84}$\textbf{Se} & \multicolumn{3}{c}{Shell 7} \\
\hdashline
$2.827 \times 10^{28}$ & $2.679 \times 10^{10}$ & 6.1 & $^{84}$\textbf{Se} + 2e$^-$ $\to$ $^{84}$\textbf{Ge} + 2$\nu_e$ & 28.6 & $1.934 \times 10^{28}$ & 216 \\ 
$1.189 \times 10^{29}$ & $8.353 \times 10^{10}$ & 6.5 & $^{84}$\textbf{Ge} + 2e$^-$ $\to$ $^{84}$Zn + 2$\nu_e$ & 33.1 & $1.100 \times 10^{29}$ & $1.2 \times 10^{3}$ \\ 
$3.255 \times 10^{29}$ & $1.895 \times 10^{11}$ & 7.0 & $^{84}$Zn + 2e$^-$ $\to$ $^{84}$Ni + 2$\nu_e$ & 29.2 & $3.166 \times 10^{29}$ & $3.5 \times 10^{3}$ \\ 
\textcolor{blue}{$7.167 \times 10^{29}$} & \textcolor{blue}{$3.668 \times 10^{11}$} & \textcolor{blue}{5.0} & \textcolor{blue}{$^{84}$Ni + 2 e$^-$ $\to$ $^{82}$Fe +  2 n + 2 $\nu_e$} & \textcolor{blue}{18.9} & \textcolor{blue}{$7.078 \times 10^{29}$} & \textcolor{blue}{$7.9 \times 10^{3}$} \\
$7.690 \times 10^{29}$ & $3.867 \times 10^{11}$ & 7.5 & $^{84}$Ni + 2e$^-$ $\to$ $^{84}$Fe + 2$\nu_e$ & 29.1 & $7.601 \times 10^{29}$ & $8.5 \times 10^{3}$ \\ 
$1.398 \times 10^{30}$ & $6.514 \times 10^{11}$ & 17.0 & $^{84}$Fe + 4e$^-$ $\to$ $^{72}$Ti + 12 n + 4$\nu_e$ & 64.8 & $1.389 \times 10^{30}$ & $1.6 \times 10^{4}$ \\ 
\hline
\end{tabular}
\end{table*}

\begin{table*}
\caption{}
\label{tab:pacoc2}
\centering
\begin{tabular}{ccccccc}
\hline 
\hline 
P(\;dyn/cm$^2$) & $\rho_{ini}$ & $\lambda$ & Reaction & Q (keV/A) & $\Delta$ P & $\delta P/P $ (in \%) \\
$8.954 \times 10^{27}$ & $1.172 \times 10^{10}$ & - & $^{82}$\textbf{Ge} & \multicolumn{3}{c}{Shell 8} \\
\hdashline
$6.325 \times 10^{28}$ & $5.077 \times 10^{10}$ & 6.5 & $^{82}$\textbf{Ge} + 2e$^-$ $\to$ $^{82}$\textbf{Zn} + 2$\nu_e$ & 23.0 & $3.488 \times 10^{28}$ & 122 \\ 
$2.991 \times 10^{29}$ & $1.736 \times 10^{11}$ & 7.0 & $^{82}$\textbf{Zn} + 2e$^-$ $\to$ $^{82}$Ni + 2$\nu_e$ & 44.3 & $2.708 \times 10^{29}$ & 954 \\ 
$5.970 \times 10^{29}$ & $3.121 \times 10^{11}$ & 7.5 & $^{82}$Ni + 2e$^-$ $\to$ $^{82}$Fe + 2$\nu_e$ & 29.9 & $5.686 \times 10^{29}$ & $2.0 \times 10^{3}$ \\ 
\textcolor{blue}{$1.032 \times 10^{30}$} & \textcolor{blue}{$5.065 \times 10^{11}$} & \textcolor{blue}{2.9} & \textcolor{blue}{$^{82}$Fe + 2 e$^-$ $\to$ $^{78}$Cr +  4 n + 2 $\nu_e$} & \textcolor{blue}{25.4} & \textcolor{blue}{$1.004 \times 10^{30}$} & \textcolor{blue}{$3.5 \times 10^{3}$} \\
$1.201 \times 10^{30}$ & $5.676 \times 10^{11}$ & 8.0 & $^{82}$Fe + 2e$^-$ $\to$ $^{78}$Cr + 4 n + 2$\nu_e$ & 22.1 & $1.173 \times 10^{30}$ & $4.1 \times 10^{3}$ \\ 
\hline
$2.846 \times 10^{28}$ & $2.897 \times 10^{10}$ & - & $^{80}$\textbf{Zn} & \multicolumn{3}{c}{Shell 9} \\
\hdashline
$2.529 \times 10^{29}$ & $1.493 \times 10^{11}$ & 7.0 & $^{80}$\textbf{Zn} + 2e$^-$ $\to$ $^{80}$Ni + 2$\nu_e$ & 56.3 & $1.597 \times 10^{29}$ & 171 \\ 
$4.503 \times 10^{29}$ & $2.464 \times 10^{11}$ & 7.5 & $^{80}$Ni + 2e$^-$ $\to$ $^{80}$Fe + 2$\nu_e$ & 30.8 & $3.571 \times 10^{29}$ & 383 \\ 
\textcolor{blue}{$8.818 \times 10^{29}$} & \textcolor{blue}{$4.389 \times 10^{11}$} & \textcolor{blue}{5.5} & \textcolor{blue}{$^{80}$Fe + 2 e$^-$ $\to$ $^{78}$Cr +  2 n + 2 $\nu_e$} & \textcolor{blue}{16.1} & \textcolor{blue}{$7.886 \times 10^{29}$} & \textcolor{blue}{846} \\
$1.022 \times 10^{30}$ & $4.902 \times 10^{11}$ & 8.2 & $^{80}$Fe + 2e$^-$ $\to$ $^{80}$Cr + 2$\nu_e$ & 29.5 & $9.284 \times 10^{29}$ & 996 \\ 
$1.500 \times 10^{30}$ & $7.076 \times 10^{11}$ & 18.4 & $^{80}$Cr + 4e$^-$ $\to$ $^{66}$Ca + 14 n + 4$\nu_e$ & 80.9 & $1.407 \times 10^{30}$ & $1.5 \times 10^{3}$ \\ 
\hline 
$9.346 \times 10^{28}$ & $7.248 \times 10^{10}$ & - & $^{82}$\textbf{Zn} & \multicolumn{3}{c}{Shell 10} \\
\hdashline
$2.991 \times 10^{29}$ & $1.736 \times 10^{11}$ & 7.0 & $^{82}$\textbf{Zn} + 2e$^-$ $\to$ $^{82}$Ni + 2$\nu_e$ & 44.3 & $1.954 \times 10^{29}$ & 188 \\ 
$5.970 \times 10^{29}$ & $3.121 \times 10^{11}$ & 7.5 & $^{82}$Ni + 2e$^-$ $\to$ $^{82}$Fe + 2$\nu_e$ & 29.9 & $4.932 \times 10^{29}$ & 475 \\ 
\textcolor{blue}{$1.032 \times 10^{30}$} & \textcolor{blue}{$5.065 \times 10^{11}$} & \textcolor{blue}{2.9} & \textcolor{blue}{$^{82}$Fe + 2 e$^-$ $\to$ $^{78}$Cr +  4 n + 2 $\nu_e$} & \textcolor{blue}{25.4} & \textcolor{blue}{$9.287 \times 10^{29}$} & \textcolor{blue}{894} \\
$1.201 \times 10^{30}$ & $5.676 \times 10^{11}$ & 8.0 & $^{82}$Fe + 2e$^-$ $\to$ $^{78}$Cr + 4 n + 2$\nu_e$ & 22.1 & $1.098 \times 10^{30}$ & $1.1 \times 10^{3}$ \\ 
\hline
$1.041 \times 10^{29}$ & $7.973 \times 10^{10}$ & - & $^{94}$Se & \multicolumn{3}{c}{Shell 11} \\
\hdashline
$2.337 \times 10^{29}$ & $1.463 \times 10^{11}$ & 6.1 & $^{94}$Se + 2e$^-$ $\to$ $^{94}$Ge + 2$\nu_e$ & 25.5 & $1.261 \times 10^{29}$ & 117 \\ 
$5.492 \times 10^{29}$ & $2.949 \times 10^{11}$ & 6.5 & $^{94}$Ge + 2e$^-$ $\to$ $^{94}$Zn + 2$\nu_e$ & 25.4 & $4.416 \times 10^{29}$ & 410 \\ 
\textcolor{blue}{$9.293 \times 10^{29}$} & \textcolor{blue}{$4.663 \times 10^{11}$} & \textcolor{blue}{2.4} & \textcolor{blue}{$^{94}$Zn + 2 e$^-$ $\to$ $^{90}$Ni +  4 n + 2 $\nu_e$} & \textcolor{blue}{16.8} & \textcolor{blue}{$8.217 \times 10^{29}$} & \textcolor{blue}{763} \\
$1.084 \times 10^{30}$ & $5.234 \times 10^{11}$ & 7.0 & $^{94}$Zn + 2e$^-$ $\to$ $^{94}$Ni + 2$\nu_e$ & 22.1 & $9.760 \times 10^{29}$ & 907 \\ 
$1.372 \times 10^{30}$ & $6.684 \times 10^{11}$ & 15.4 & $^{94}$Ni + 4e$^-$ $\to$ $^{80}$Cr + 14 n + 4$\nu_e$ & 54.7 & $1.264 \times 10^{30}$ & $1.2 \times 10^{3}$ \\ 
\hline
$1.079 \times 10^{29}$ & $8.242 \times 10^{10}$ & - & $^{100}$Kr & \multicolumn{3}{c}{Shell 12} \\
\hdashline
$2.316 \times 10^{29}$ & $1.462 \times 10^{11}$ & 5.7 & $^{100}$Kr + 2e$^-$ $\to$ $^{100}$Se + 2$\nu_e$ & 23.7 & $1.051 \times 10^{29}$ & 83 \\ 
$5.256 \times 10^{29}$ & $2.861 \times 10^{11}$ & 6.1 & $^{100}$Se + 2e$^-$ $\to$ $^{100}$Ge + 2$\nu_e$ & 23.6 & $3.992 \times 10^{29}$ & 315 \\ 
\textcolor{blue}{$8.827 \times 10^{29}$} & \textcolor{blue}{$4.481 \times 10^{11}$} & \textcolor{blue}{2.3} & \textcolor{blue}{$^{100}$Ge + 2 e$^-$ $\to$ $^{96}$Zn +  4 n + 2 $\nu_e$} & \textcolor{blue}{13.6} & \textcolor{blue}{$7.562 \times 10^{29}$} & \textcolor{blue}{597} \\
$1.030 \times 10^{30}$ & $5.034 \times 10^{11}$ & 6.5 & $^{100}$Ge + 2e$^-$ $\to$ $^{100}$Zn + 2$\nu_e$ & 22.7 & $9.039 \times 10^{29}$ & 714 \\ 
$1.382 \times 10^{30}$ & $6.683 \times 10^{11}$ & 14.3 & $^{100}$Zn + 4e$^-$ $\to$ $^{86}$Fe + 14 n + 4$\nu_e$ & 57.7 & $1.255 \times 10^{30}$ & 992 \\ 
\hline
$1.269 \times 10^{29}$ & $9.446 \times 10^{10}$ & - & $^{96}$Se & \multicolumn{3}{c}{Shell 13} \\
\hdashline
$3.149 \times 10^{29}$ & $1.869 \times 10^{11}$ & 6.1 & $^{96}$Se + 2e$^-$ $\to$ $^{96}$Ge + 2$\nu_e$ & 24.8 & $1.845 \times 10^{29}$ & 141 \\ 
\textcolor{blue}{$6.630 \times 10^{29}$} & \textcolor{blue}{$3.470 \times 10^{11}$} & \textcolor{blue}{4.3} & \textcolor{blue}{$^{96}$Ge + 2 e$^-$ $\to$ $^{94}$Zn +  2 n + 2 $\nu_e$} & \textcolor{blue}{19.5} & \textcolor{blue}{$5.327 \times 10^{29}$} & \textcolor{blue}{408} \\
$6.921 \times 10^{29}$ & $3.584 \times 10^{11}$ & 6.5 & $^{96}$Ge + 2e$^-$ $\to$ $^{96}$Zn + 2$\nu_e$ & 24.8 & $5.618 \times 10^{29}$ & 431 \\ 
$1.246 \times 10^{30}$ & $5.935 \times 10^{11}$ & 6.9 & $^{96}$Zn + 2e$^-$ $\to$ $^{92}$Ni + 4 n + 2$\nu_e$ & 22.0 & $1.115 \times 10^{30}$ & 855 \\ 
\hline
$1.307 \times 10^{29}$ & $9.709 \times 10^{10}$ & - & $^{102}$Kr & \multicolumn{3}{c}{Shell 14} \\
\hdashline
$3.076 \times 10^{29}$ & $1.846 \times 10^{11}$ & 5.7 & $^{102}$Kr + 2e$^-$ $\to$ $^{102}$Se + 2$\nu_e$ & 23.1 & $1.322 \times 10^{29}$ & 75 \\ 
\textcolor{blue}{$6.369 \times 10^{29}$} & \textcolor{blue}{$3.372 \times 10^{11}$} & \textcolor{blue}{4.0} & \textcolor{blue}{$^{102}$Se + 2 e$^-$ $\to$ $^{100}$Ge +  2 n + 2 $\nu_e$} & \textcolor{blue}{19.6} & \textcolor{blue}{$4.616 \times 10^{29}$} & \textcolor{blue}{263} \\
$6.566 \times 10^{29}$ & $3.449 \times 10^{11}$ & 6.1 & $^{102}$Se + 2e$^-$ $\to$ $^{102}$Ge + 2$\nu_e$ & 23.0 & $4.813 \times 10^{29}$ & 274 \\ 
$1.178 \times 10^{30}$ & $5.679 \times 10^{11}$ & 6.4 & $^{102}$Ge + 2e$^-$ $\to$ $^{98}$Zn + 4 n + 2$\nu_e$ & 18.5 & $1.003 \times 10^{30}$ & 572 \\ 
\hline
$1.759 \times 10^{29}$ & $1.237 \times 10^{11}$ & - & $^{104}$Kr & \multicolumn{3}{c}{Shell 15} \\
\hdashline
$3.976 \times 10^{29}$ & $2.282 \times 10^{11}$ & 5.7 & $^{104}$Kr + 2e$^-$ $\to$ $^{104}$Se + 2$\nu_e$ & 22.5 & $1.646 \times 10^{29}$ & 70 \\ 
\textcolor{blue}{$7.340 \times 10^{29}$} & \textcolor{blue}{$3.824 \times 10^{11}$} & \textcolor{blue}{4.1} & \textcolor{blue}{$^{104}$Se + 2 e$^-$ $\to$ $^{102}$Ge +  2 n + 2 $\nu_e$} & \textcolor{blue}{11.8} & \textcolor{blue}{$5.009 \times 10^{29}$} & \textcolor{blue}{214} \\
$8.051 \times 10^{29}$ & $4.099 \times 10^{11}$ & 6.1 & $^{104}$Se + 2e$^-$ $\to$ $^{104}$Ge + 2$\nu_e$ & 22.5 & $5.720 \times 10^{29}$ & 245 \\ 
$1.338 \times 10^{30}$ & $6.369 \times 10^{11}$ & 21.3 & $^{104}$Ge + 6e$^-$ $\to$ $^{86}$Fe + 18 n + 6$\nu_e$ & 75.9 & $1.105 \times 10^{30}$ & 473 \\ 
\hline
$2.337 \times 10^{29}$ & $1.561 \times 10^{11}$ & - & $^{106}$Kr & \multicolumn{3}{c}{Shell 16} \\
\hdashline
$5.025 \times 10^{29}$ & $2.774 \times 10^{11}$ & 5.7 & $^{106}$Kr + 2e$^-$ $\to$ $^{106}$Se + 2$\nu_e$ & 22.0 & $2.037 \times 10^{29}$ & 68 \\ 
\textcolor{blue}{$8.389 \times 10^{29}$} & \textcolor{blue}{$4.309 \times 10^{11}$} & \textcolor{blue}{4.1} & \textcolor{blue}{$^{106}$Se + 2 e$^-$ $\to$ $^{104}$Ge +  2 n + 2 $\nu_e$} & \textcolor{blue}{11.2} & \textcolor{blue}{$5.401 \times 10^{29}$} & \textcolor{blue}{180} \\
$9.713 \times 10^{29}$ & $4.811 \times 10^{11}$ & 6.1 & $^{106}$Se + 2e$^-$ $\to$ $^{106}$Ge + 2$\nu_e$ & 22.0 & $6.724 \times 10^{29}$ & 225 \\ 
$1.385 \times 10^{30}$ & $6.665 \times 10^{11}$ & 21.1 & $^{106}$Ge + 6e$^-$ $\to$ $^{86}$Fe + 20 n + 6$\nu_e$ & 83.6 & $1.086 \times 10^{30}$ & 363 \\ 
\hline
$2.997 \times 10^{29}$ & $1.886 \times 10^{11}$ & - & $^{112}$Sr & \multicolumn{3}{c}{Shell 17} \\
\hdashline
$4.799 \times 10^{29}$ & $2.686 \times 10^{11}$ & 5.4 & $^{112}$Sr + 2e$^-$ $\to$ $^{112}$Kr + 2$\nu_e$ & 20.5 & $1.784 \times 10^{29}$ & 59 \\ 
\textcolor{blue}{$7.977 \times 10^{29}$} & \textcolor{blue}{$4.147 \times 10^{11}$} & \textcolor{blue}{3.8} & \textcolor{blue}{$^{112}$Kr + 2 e$^-$ $\to$ $^{110}$Se +  2 n + 2 $\nu_e$} & \textcolor{blue}{10.4} & \textcolor{blue}{$4.962 \times 10^{29}$} & \textcolor{blue}{164} \\
$9.103 \times 10^{29}$ & $4.579 \times 10^{11}$ & 5.7 & $^{112}$Kr + 2e$^-$ $\to$ $^{112}$Se + 2$\nu_e$ & 20.6 & $6.087 \times 10^{29}$ & 201 \\ 
$1.344 \times 10^{30}$ & $6.490 \times 10^{11}$ & 19.6 & $^{112}$Se + 6e$^-$ $\to$ $^{92}$Ni + 20 n + 6$\nu_e$ & 76.6 & $1.043 \times 10^{30}$ & 345 \\ 
\hline

\hline
\end{tabular}
\end{table*}

\begin{table*}
\caption{}
\label{tab:pacoc3}
\centering
\begin{tabular}{ccccccc}
\hline 
\hline 
P(\;dyn/cm$^2$) & $\rho_{ini}$ & $\lambda$ & Reaction & Q (keV/A) & $\Delta$ P & $\delta P/P $ (in \%) \\
$3.024 \times 10^{29}$ & $1.933 \times 10^{11}$ & - & $^{114}$Sr & \multicolumn{3}{c}{Shell 18} \\
\hdashline
\textcolor{blue}{$5.865 \times 10^{29}$} & \textcolor{blue}{$3.178 \times 10^{11}$} & \textcolor{blue}{3.6} & \textcolor{blue}{$^{114}$Sr + 2 e$^-$ $\to$ $^{112}$Kr +  2 n + 2 $\nu_e$} & \textcolor{blue}{19.4} & \textcolor{blue}{$2.091 \times 10^{29}$} & \textcolor{blue}{55} \\
$5.906 \times 10^{29}$ & $3.195 \times 10^{11}$ & 5.4 & $^{114}$Sr + 2e$^-$ $\to$ $^{114}$Kr + 2$\nu_e$ & 20.1 & $2.131 \times 10^{29}$ & 56 \\ 
$1.058 \times 10^{30}$ & $5.218 \times 10^{11}$ & 5.7 & $^{114}$Kr + 2e$^-$ $\to$ $^{114}$Se + 2$\nu_e$ & 18.0 & $6.806 \times 10^{29}$ & 180 \\ 
$1.270 \times 10^{30}$ & $6.330 \times 10^{11}$ & 12.3 & $^{114}$Se + 4e$^-$ $\to$ $^{98}$Zn + 16 n + 4$\nu_e$ & 42.2 & $8.928 \times 10^{29}$ & 236 \\ 
\hline
$3.786 \times 10^{29}$ & $2.328 \times 10^{11}$ & - & $^{116}$Sr & \multicolumn{3}{c}{Shell 19} \\
\hdashline
\textcolor{blue}{$6.694 \times 10^{29}$} & \textcolor{blue}{$3.572 \times 10^{11}$} & \textcolor{blue}{3.6} & \textcolor{blue}{$^{116}$Sr + 2 e$^-$ $\to$ $^{114}$Kr +  2 n + 2 $\nu_e$} & \textcolor{blue}{13.0} & \textcolor{blue}{$2.039 \times 10^{29}$} & \textcolor{blue}{43} \\
$7.152 \times 10^{29}$ & $3.753 \times 10^{11}$ & 5.4 & $^{116}$Sr + 2e$^-$ $\to$ $^{116}$Kr + 2$\nu_e$ & 19.7 & $2.496 \times 10^{29}$ & 53 \\ 
$1.192 \times 10^{30}$ & $5.807 \times 10^{11}$ & 11.9 & $^{116}$Kr + 4e$^-$ $\to$ $^{106}$Ge + 10 n + 4$\nu_e$ & 35.9 & $7.265 \times 10^{29}$ & 156 \\ 
\hline
$4.669 \times 10^{29}$ & $2.772 \times 10^{11}$ & - & $^{118}$Sr & \multicolumn{3}{c}{Shell 20} \\
\hdashline
\textcolor{blue}{$7.587 \times 10^{29}$} & \textcolor{blue}{$3.992 \times 10^{11}$} & \textcolor{blue}{3.6} & \textcolor{blue}{$^{118}$Sr + 2 e$^-$ $\to$ $^{116}$Kr +  2 n + 2 $\nu_e$} & \textcolor{blue}{9.7} & \textcolor{blue}{$1.948 \times 10^{29}$} & \textcolor{blue}{34} \\
$8.540 \times 10^{29}$ & $4.362 \times 10^{11}$ & 5.4 & $^{118}$Sr + 2e$^-$ $\to$ $^{118}$Kr + 2$\nu_e$ & 19.3 & $2.901 \times 10^{29}$ & 51 \\ 
$1.305 \times 10^{30}$ & $6.322 \times 10^{11}$ & 18.5 & $^{118}$Kr + 6e$^-$ $\to$ $^{100}$Zn + 18 n + 6$\nu_e$ & 70.9 & $7.407 \times 10^{29}$ & 131 \\ 
\hline
$5.656 \times 10^{29}$ & $3.256 \times 10^{11}$ & - & $^{120}$Sr & \multicolumn{3}{c}{Shell 21} \\
\hdashline
\textcolor{blue}{$8.543 \times 10^{29}$} & \textcolor{blue}{$4.438 \times 10^{11}$} & \textcolor{blue}{1.9} & \textcolor{blue}{$^{120}$Sr + 2 e$^-$ $\to$ $^{116}$Kr +  4 n + 2 $\nu_e$} & \textcolor{blue}{12.1} & \textcolor{blue}{$1.953 \times 10^{29}$} & \textcolor{blue}{29} \\
$1.004 \times 10^{30}$ & $5.009 \times 10^{11}$ & 5.4 & $^{120}$Sr + 2e$^-$ $\to$ $^{120}$Kr + 2$\nu_e$ & 18.5 & $3.448 \times 10^{29}$ & 52 \\ 
$1.272 \times 10^{30}$ & $6.310 \times 10^{11}$ & 18.4 & $^{120}$Kr + 6e$^-$ $\to$ $^{100}$Zn + 20 n + 6$\nu_e$ & 61.2 & $6.134 \times 10^{29}$ & 93 \\ 
\hline
$6.609 \times 10^{29}$ & $3.714 \times 10^{11}$ & - & $^{128}$Zr & \multicolumn{3}{c}{Shell 22} \\
\hdashline
\textcolor{blue}{$8.955 \times 10^{29}$} & \textcolor{blue}{$4.666 \times 10^{11}$} & \textcolor{blue}{0.2} & \textcolor{blue}{$^{128}$Zr + 2 e$^-$ $\to$ $^{122}$Sr +  6 n + 2 $\nu_e$} & \textcolor{blue}{14.7} & \textcolor{blue}{$1.324 \times 10^{29}$} & \textcolor{blue}{17} \\
$1.067 \times 10^{30}$ & $5.321 \times 10^{11}$ & 5.1 & $^{128}$Zr + 2e$^-$ $\to$ $^{128}$Sr + 2$\nu_e$ & 14.8 & $3.036 \times 10^{29}$ & 39 \\ 
$1.150 \times 10^{30}$ & $5.919 \times 10^{11}$ & 5.2 & $^{128}$Sr + 2e$^-$ $\to$ $^{118}$Kr + 10 n + 2$\nu_e$ & 16.5 & $3.871 \times 10^{29}$ & 50 \\ 
\hline
$7.654 \times 10^{29}$ & $4.212 \times 10^{11}$ & - & $^{130}$Zr & \multicolumn{3}{c}{Shell 23} \\
\hdashline
\textcolor{blue}{$8.969 \times 10^{29}$} & \textcolor{blue}{$4.744 \times 10^{11}$} & \textcolor{blue}{-1.4} & \textcolor{blue}{$^{130}$Zr + 2 e$^-$ $\to$ $^{122}$Sr +  8 n + 2 $\nu_e$} & \textcolor{blue}{14.6} & \textcolor{blue}{$9.688 \times 10^{27}$} & \textcolor{blue}{1.092} \\
$1.188 \times 10^{30}$ & $5.860 \times 10^{11}$ & 16.4 & $^{130}$Zr + 6e$^-$ $\to$ $^{112}$Se + 18 n + 6$\nu_e$ & 54.0 & $3.009 \times 10^{29}$ & 33 \\ 
\hline
$8.895 \times 10^{29}$ & $4.787 \times 10^{11}$ & - & $^{132}$Zr & \multicolumn{3}{c}{Shell 24} \\
\hdashline
$1.248 \times 10^{30}$ & $6.174 \times 10^{11}$ & 16.2 & $^{132}$Zr + 6e$^-$ $\to$ $^{112}$Se + 20 n + 6$\nu_e$ & 64.4 & $3.047 \times 10^{29}$ & 32 \\ 
\hline
\end{tabular}
\end{table*}

\end{appendix}

\end{document}